\documentclass[aps,reprint]{revtex4-1}
\usepackage{blindtext}

\usepackage{amsfonts}
\usepackage{color}    
\usepackage{graphicx}
\usepackage{braket} 
\usepackage{dcolumn}
\usepackage{bm}
\usepackage{subfigure}
\usepackage{amssymb}
\usepackage{multirow}
\usepackage{natbib}
\usepackage{amsmath}
\usepackage{mathtools}
\usepackage{xcolor}

\newcommand{\db}[2][]{\text{d}^{#1}#2}

\newcommand{\sgn}{\text{\normalfont Sgn}}

\DeclarePairedDelimiter\abs{\lvert}{\rvert}

\DeclarePairedDelimiter\Avr{\langle}{\rangle}

\begin{document}

\title{Analysis of dynamical effects in the uniform electron liquids with the self-consistent method of moments complemented by the Shannon information entropy and the path-integral Monte-Carlo simulations}


\author{A.V. Filinov}
\email{filinov@theo-physik.uni-kiel.de}
\affiliation{Institut für Theoretische Physik und Astrophysik, Christian-Albrechts-Universität zu Kiel, Kiel, Germany}

\author{J. Ara}
\affiliation{Instituto de Tecnolog\'{\i}a Qu\'{\i}mica,
Universitat Polit\`{e}cnica de Val\`{e}ncia-Consejo Superior de Investigaciones Cient\'{\i}ficas, Valencia, Spain}

\author{I.M. Tkachenko}
\email{imtk@mat.upv.es}
\affiliation{Departament de Matem\`{a}tica Aplicada, Universitat Poli\`{e}cnica de Val\`{e}ncia, Valencia, Spain}
\affiliation{Al-Farabi Kazakh National University, Almaty, Kazakhstan}

\date{20.03.2023}

\begin{abstract}
Dynamical properties of uniform electron fluids are studied within a non-perturbative approach consisting in the combination of the self-consistent version of the method of moments (SCMM) involving up to nine sum rules and other exact relations, the two-parameter Shannon information entropy maximization procedure, and the $\textit{ab initio}$ path integral Monte Carlo (PIMC) simulations of the imaginary-time intermediate scattering function.

The explicit dependence of the dynamic structure factor (DSF) on temperature and density is studied in a broad realm of variation of the dimensionless parameters ($2\leq r_s\leq 36$ and $1\leq \theta \leq 8$). When the coupling is strong ($r_s\geq 16$) we clearly observe a bi-modal structure of the excitation spectrum with a lower-energy mode possessing a well pronounced roton-like feature ($\theta \leq 2$) and an additional high-energy branch within the roton region which evolves into the strongly overdamped high-frequency shoulder when the coupling decreases ($r_s\leq 10$). We are not aware of any reconstruction of the DSF at these
conditions with the effects of dynamical correlations included here via the intermediate scattering and the dynamical Nevanlinna parameter functions. The standard static-local-field approach fails to reproduce this effect. The reliability of our method is confirmed by a detailed comparison with the recent $\textit{ab initio}$ dynamic-local-field approach by Dornheim {\it et al.} [Phys.Rev.Lett. {\bf 121}, 255001 (2018)] available for high/moderate densities ($r_s\leq 10$).
Moreover, within the SCMM we are able to construct the modes’ dispersion equation in a closed analytical form and find the decrements (lifetimes) of the quasiparticle excitations explicitly. The physical nature of the revealed modes is discussed. 

Mathematical details of the method are complemented in the Appendix. The proposed approach, due to its rigorous mathematical foundation, can find numerous diverse applications in the physics of Fermi and Bose liquids.

\end{abstract}

\maketitle

\section{Introduction.} 
Fermionic and bosonic three- and
two-dimensional fluids of charged or neutral particles (see ~\cite{book_gases,book_plasma,filinov.2012pra,PhysRevA.87.033624,filinov.2016pra,dornheim.2021cpp,Philos.Trans.RoyalSoc.A} and references therein) constitute an important class of one-component systems which serve to testify different theoretical models and are of significant practical
importance for the interpretation and development of real experimental studies~\cite{Thomson_scattering,Ionization_potential}. In this list, uniform electron fluids and, in particular, the uniform electron gas (UEG), an exotic, highly compressed neutral Coulomb system between solid and plasma phases~\cite{dornheim_physrep_18}, is one of the key models of the warm dense matter (WDM). 
This model system is of importance for our understanding of planet interiors~\cite{PhysRevLett.108.091102,Militzer_2008}, laser excited
solids~\cite{Ernstorfer_2009} or inertial confinement fusion~\cite{PhysRevLett.114.045001,Schmit.prl.2014,Nature.Physics.2016,PhysRevLett.109.225001}.

The most accurate results in the WDM regime, so far, have been obtained via the first-principles methods of numerical simulations such as the quantum Monte Carlo (QMC)~\cite{Brown.prl2013,Schoof.2015,Malone.jcp2015,Malone.prl2016,dornheim.prb2016,PhysRevE.102.033203,filinov-etal.2021ctpp,qmc_pot2022,qmc_spin2022}.
Despite quite accurate results for the static properties, the extraction of a similar quality QMC data for the dynamical characteristics (dynamic conductivity, optical absorption, collective excitations) is quite difficult and until recently~\cite{dornheim-etal.nl.2020prl} has been realized within the linear response theory. In particular, the dynamic structure factor (DSF), $S(k,\omega),$ is the central quantity in the x-ray Thomson scattering diagnostics of the WDM realized nowadays at large research facilities~\cite{NIF2010,app7060592,xray2015}. QMC simulations do not provide direct access to this quantity but permit to obtain reliable results with respect to the intermediate scattering function  
\begin{eqnarray}
F(\mathbf{q},\tau)=\Avr{\rho_{\mathbf{q}}(0)\rho_{\mathbf{-q}}(\tau)}=\int\limits_{-\infty}^{\infty} S(\mathbf{q},\omega)\, e^{-\hbar \tau \omega}\, \db \omega\
\label{Fqw}
\end{eqnarray}
computed in the "imaginary" time $\hbar \tau \in [0,\hbar \beta]$, $\beta$ being the inverse temperature in energy units.

The inversion of the Laplace transform~(\ref{Fqw}) for $S(q,\omega)$ can be realized via the maximum entropy method~\cite{silver.1990prb} or the stochastic and the generic optimizations~\cite{mishchenko.prb2000,vitali.2010prb,filinov.2012pra}. 
Such a reconstruction is, unfortunately, not unique and the space of trial solutions expands with the increase of the statistical noise in $F(q,\tau)$. Nevertheless, a number of trial solutions can be drastically reduced using a set of restrictions imposed either by several frequency moments of the spectral density~\cite{filinov.2016pra} or relying on the exact properties of the dynamic local field correction (DLFC). This permits to reconstruct the most accurate UEG DSF~\cite{dornheim.2018prl,groth.2019prb}. The stochastic sampling of the trial solutions for the DLFC, however, 
is computationally expensive. Moreover, an accurate estimation of $F(q,\tau)$ requires time-consuming  simulations and is limited to the temperature-density realm where QMC is not disabled by the fermion sign problem~\cite{troyer.2005prl}. In addition, the lower boundary of accessible wavenumbers is limited by the system size, i.e. $q\geq 2 \pi /L$ with  $L=(N/n)^{1/3}$. The algorithmic Matsubara diagrammatic Monte Carlo technique seems to be even more computationally involved~\cite{diagrams_2022}. 

As an alternative to the DLFC-based reconstruction~\cite{dornheim.2018prl,groth.2019prb} with much lower computational demand and applicability to a much broader class of physical systems, we present here the {\it nine}-moment version of the original non-perturbative self-consistent~\cite{arkhipov-etal.2017prl,arkhipov-etal.2020pre} method of moments~\cite{nevanlinna-book,shohat-book,krein-book,akhiezer-book,tkachenko-book} complemented by the Shannon information entropy two-parameter maximization technique and other exact requirements. Within this approach the DSF sum rules known theoretically or numerically are incorporated into the analytical form of the spectral density automatically. 
Resulting DSF $S\left( q,\omega\right)$, the (inverse) longitudinal dielectric function $\epsilon^{-1}\left( q,\omega\right)$, the eigenmode spectrum and other dynamical characteristics are constructed exclusively in terms of the static structure factor (SSF), $S\left(q\right)=F\left( q,0\right)$, and the static dielectric function,  $\epsilon\left( q,0\right)$. These input data are provided here by the recent fermionic QMC simulations~\cite{filinov-etal.2021ctpp}.

The respective accuracy of our approach is demonstrated and opens a path to further improvements and extensions to a broader parameter domain. A simplified version of our method was also validated against the QMC static data~\cite{ara-etal.2021pop}. Preliminary steps to the creation of the present approach were taken in~\cite{PNP, Ara_2022}.   

In what follows we use the reduced temperature, $\theta=k_B T/E_F$ with $E_F=(\hbar^2/2m) (3 \pi^2 n)^{2/3}$, and the density (Brueckner) parameter $r_s$ defined by $n a_B^3=(4\pi r_s^3/3)^{-1}$.
Here, $n$\ is the number density of charged particles, $a_B$ is the first Bohr radius, and $E_F$ is the Fermi energy.

In the present paper we concentrate first on the warm dense matter regime~\cite{dornheim_physrep_18} with the coupling ($r_s$) and degeneracy ($\theta$) parameters varying around unity ($\theta, r_s\sim 1$). Then we extend our studies to the strongly coupled regime defined by $10 \leq r_{s} \leq 36$. 

The paper is organized as follows. In Sec.~\ref{pimcSec} we describe some details of the performed QMC simulations: we briefly mention the manifestations of the fermion sign problem in our simulations, demonstrate the convergence of main thermodynamic properties and the influence of the finite-size effects.
The generalized self-consistent method of moments (SCMM) with the dynamical Nevanlinna function is presented and discussed in detail in Sec.~\ref{SCMM}. The UEG eigenmodes and the dynamic structure factor at moderate densities ($2\leq r_s\leq 10$) are obtained and compared to the local-field-based data.
Further, in Sec.~\ref{Corr_effects} we present an improved version of the method. It is based on the optimization of the dynamical Nevanlinna function with an additional information contained in the intermediate scattering function $F(q,\tau)$ provided by ab initio path-integral Monte-Carlo (PIMC) simulations. This new combined approach allows to study the influence of multiple correlation effects on the dynamical response in the UEG in the low density phase ($16\leq r_s\leq 36$) for the first time. Main conclusions and the outlook are drawn in Sec.~\ref{Outlook}.  

\section{The path-integral Monte Carlo simulation method}\label{pimcSec}

\subsection{Account of Fermi-Dirac statistics in QMC simulations}

In this section we briefly introduce the fermionic propagator path integral (FP-PIMC) recently developed by Filinov {\it et al.}~\cite{filinov-etal.2021ctpp} which provides the UEG ab initio static properties which are further employed in the self-consistent method of moments (Sec.~\ref{SCMM}) to recover the dynamical response.

The FP-PIMC has demonstrated its efficiency in the analysis of the exchange correlation free energy for the UEG jellium model in a broad realm of parameters: $0.1\leq r_s\leq 10$ and $1\leq \theta\leq 2$. 

In contrast to the standard high-temperature decomposition of the fermionic partition function $\mathbb Z_F$ via the bosonic propagators~\cite{RevModPhys.67.279}, the FP-PIMC employs the anti-symmetric one, in the form of many-body Slater determinants, which already satisfy the required symmetry relations under an exchange of identical fermions. The summation over different permutation classes~\cite{RevModPhys.67.279}, $\{\sigma_s\}$, can be performed analytically in the kinetic energy part of the $N$-body density matrix. As a result the  anti-symmetric (fermionic) free-particle propagators (denoted in the following as "FP") between two adjacent time-slices are expressed as follows 
 \begin{eqnarray}
 D_{p-1,p}^s=&&\sum\limits_{\sigma_s} \left\langle \mathbf R_{p-1}|e^{- \epsilon \hat K} | \hat \pi_{\sigma_s} \mathbf R_{p} \right\rangle=\frac{1}{\lambda^{DN^s}} \det \mathbb M^s_{p-1,p}\label{M1}
\end{eqnarray}  
where $\mathbb M^s_{p-1,p}$ is the $N^s \times N^s$ diffusion matrix
\begin{eqnarray}
&&\mathbb M^s_{p-1,p} = ||m_{kl}(p-1,p)||, \quad k,l=1,\ldots N^s,\\
&&m_{kl}(p-1,p) = \exp\left(-\frac{\pi}{\lambda^2_{\epsilon}}\left[\mathbf r^s_{l\, p}-\mathbf r^s_{k\, (p-1)}\right]^2\right)   \label{M2}\,.
\end{eqnarray} 
To shorten the notations, we introduced the total radius vector for identical particles of the same type, $\mathbf R^{s}_{p}=(\mathbf r^s_ {1 \,p},\ldots,\mathbf r^s_ {N^s\,p})$, where the upper index denotes the spin state $s=\{\uparrow,\downarrow \}$, the first lower index counts the particle number indices ($1 \ldots N_s$), and the second lower index denotes the imaginary time argument, $\tau_p= p \epsilon$, with $\epsilon=\beta/ P$ and $0\leq p\leq P$. Next,
we can define the space-time variable, $\mathbf X^s=(\mathbf R^s_1,\ldots, \mathbf R^s_P)$, which specifies a system microstate -- a specific microscopic configuration of particle trajectories. 
The resulting expression for the partition function $\mathbb Z_F$ 
thus contains the Slater determinants, $\mathbb{M}^{s}_{p-1,p}$, between each successive imaginary times $\tau_{p}-\tau_{p-1}=\epsilon$, and, for practical applications in the Monte Carlo methods can be rewritten in the equivalent form with a new effective action $S_A(p-1,p)$ which along with the standard potential energy term $U$  contains an additional exchange contribution $W_{\text{x}}$
\begin{eqnarray}
&&\mathbb Z_F =\frac{1}{N^{\uparrow}! N^{\downarrow}!} \int \db \mathbf X^{\uparrow} \db \mathbf X^{\downarrow} \prod\limits_{p=1}^P \sgn_p \cdot e^{-S_A(p-1,p)},\label{ZFF}\\
&&e^{-S_A(p-1,p)} = e^{-\epsilon U(R^{\uparrow}_{p},R_{p}^{\downarrow})} \cdot  e^{W_{\text{x}}(R^{\uparrow}_{p},R_{p}^{\downarrow})},\\
&&W_{\text{x}} =\ln \abs{\det \mathbb M^{\uparrow}_{p-1,p}}+\ln \abs{\det \mathbb M^{\downarrow}_{p-1,p}}.\label{Wex}
\end{eqnarray}
Hence, the probability of microstates sampled with the new action $S_A$ becomes proportional to the absolute value of the Slater determinants. Their degeneracy in the microstates with small spatial separations of the spin-like electrons correctly recovers the Pauli blocking effect and increases the average sign $\Avr{S}$, Eq.~(\ref{def_Sign}), being crucial for the numerical accuracy of the estimated physical observables (see below). The similar idea has been employed by several authors in different  physical applications~\cite{Takahashi1984,filinov_ppcf_01,Lyubartsev_2005,Chin2015,dornheim_njp15} including the uniform electron gas at warm dense matter conditions~\cite{dornheim_physrep_18}.  

The change in the sign of Slater determinants evaluated along the imaginary time, $0\leq \tau_p\leq \beta$, is taken into account by extra factors, $\sgn_p$. Combined together they define the average sign in the fermionic PIMC, 
\begin{eqnarray}
\Avr{S}=\Avr{\prod\limits_{p=1}^P\sgn \, \mathbb{M}^{\uparrow}_{p-1,p} \cdot \sgn  \mathbb{M}^{\downarrow}_{p-1,p}},\label{def_Sign}    
\end{eqnarray}
and characterize the efficiency of simulations, as the statistical error $\delta A$ of the estimated thermodynamic observables, $\bar{A}=\Avr{A}\pm \delta A$, is scaled as 
 $\delta A \sim 1/\Avr{S(N,\beta)}$. The PIMC simulations become hampered by the fermion sign problem~\cite{CeperleyFermi,Troyer2005} once the statistical uncertainties are strongly enhanced due to an exponential decay of the average sign $\Avr{S(N,\beta)}$ with the particle number $N$, the inverse temperature $\beta=1/k_B T$ or the degeneracy parameter, $\theta=T/T_F$ (or $\chi=n \lambda^3$).
The usage of the {\it fermionic propagators}, Eq.~(\ref{M1}), permits to partially overcome the sign problem and make the UEG simulations feasible up to the degeneracy factor $n \lambda^3 \lesssim 3$ 
($\lambda$ being the themal de Broglie wavelength) 
with the average sign staying above $\Avr{S}\gtrsim 10^{-2}$, see Ref.~\cite{filinov-etal.2021ctpp}.

\subsection{High-temperature factorization and the convergence tests}

The next issue which strongly influences the efficiency of PIMC simulations is the discretization time step $\epsilon=\beta/P$. The general problem is related with the inability to estimate the exact value of the matrix elements of the density operator, $e^{-\beta \hat H}$, due to the non-commutability of the kinetic and the potential energy operators. This issue was elegantly solved by R.Feymann~\cite{Feynmanbook}, who proposed to map the original quantum partition function to a quasi-classical one at a new effective high temperature, $\tilde T=1/\epsilon=P \cdot T$, by employing the semi-group property of the evolution operator, $e^{-\beta \hat H}=\left(e^{-\epsilon \hat H}\right)^P$. This idea renders to the high-temperature factorization representation~(\ref{ZFF}).  
In the fermionic simulations the use of a larger time step $\epsilon$ (smaller $P$-value)
increases the $\Avr{S}$-value and extends the applicability range of the method to a higher degeneracy~\cite{Chin2015}. 

To reduce a number of $P$ factors in the DM we implement the fourth-order factorization scheme introduced by 
Chin \textit{et al.}~\cite{Chin2002} and Sakkos \textit{et al.}~\cite{Sakkos2009}:
\begin{eqnarray}
  &&e^{-\beta\hat H}
     =\prod\limits_{p=1}^P e^{-\epsilon(\hat K+\hat V) } \label{4thorder} \\ 
      &&\approx \prod\limits_{p=1}^P e^{-\epsilon \hat W_{1}} e^{-t_1 \epsilon \hat K}  e^{- \epsilon \hat W_{2}} e^{-t_1 \epsilon \hat K} e^{\epsilon \hat W_{1}} e^{-t_0 \epsilon \hat K} + O(\epsilon^{4})\,,
     \nonumber
\end{eqnarray}
with the choice $\epsilon=\beta/P,\,  (2t_1+t_0=1),\, t_0=1/6$, and $\hat K (\hat V)$ being the kinetic (potential) energy operator. 

\begin{figure}[t]
\begin{center}
\hspace{-0.3cm}
\includegraphics[width=0.51\textwidth]{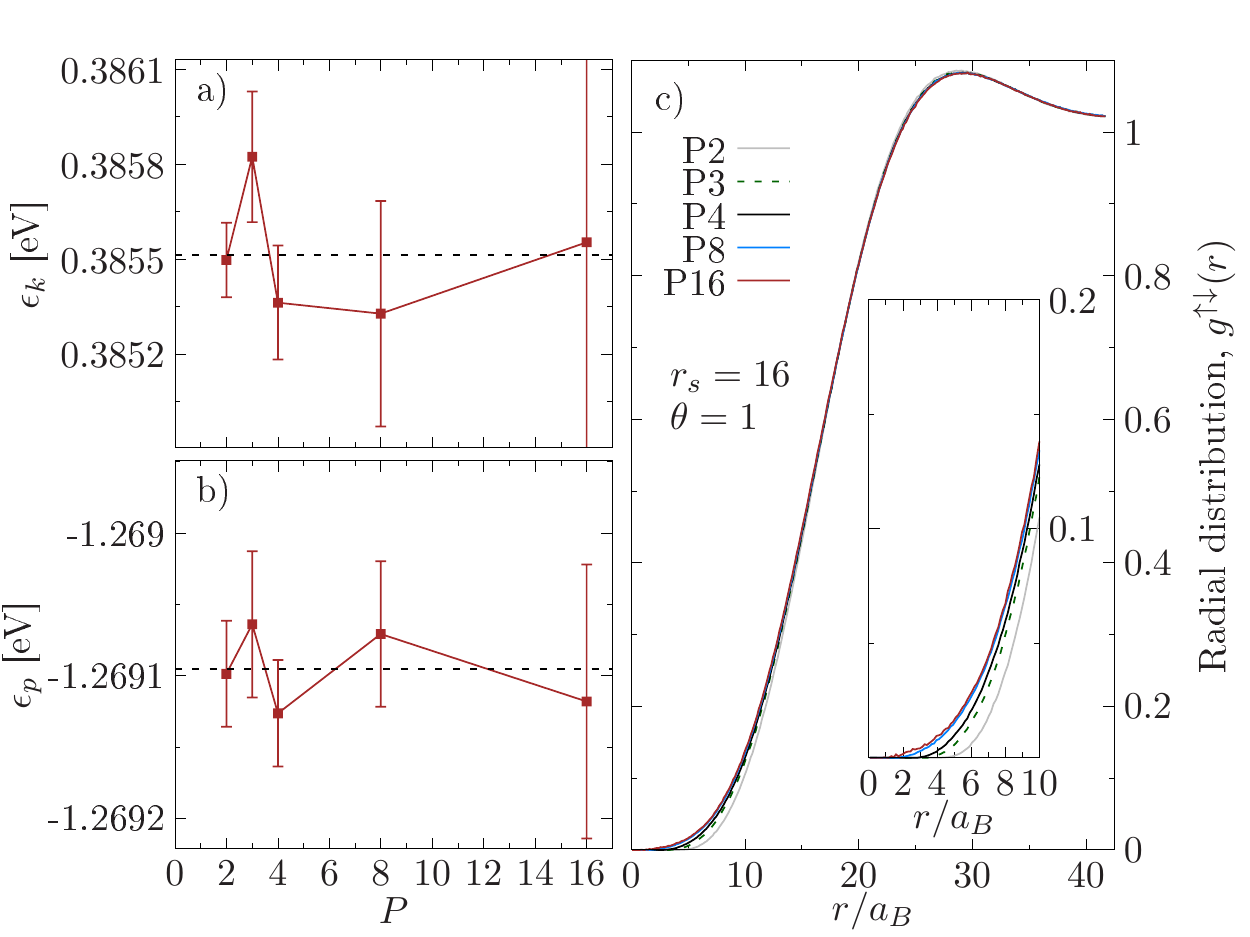}
\vspace{-0.7cm}
\caption{The $P$-convergence test for the UEG at $r_s=16$ and $\theta=1$. 
The FP-PIMC results are shown vs. the number of factorization factors $P$ in $\mathbb Z_F$, 
see Eq.~(\ref{ZFF}). Panels a,b: 
The kinetic, $\epsilon_k$, and the potential energy, $\epsilon_p$, 
per electron for $2\leq P\leq 16$. The employment of the fourth-order propagators already delivers converged results for $P=2$ (the average value extrapolated to $P\gg 1$ is shown by the dashed black line). Panel c: The corresponding $P$-convergence for the radial distribution function for the spin-unlike electrons, $g^{\uparrow \downarrow}(r)$. Some noticeable deviations are mainly observed at smaller distances, $r\lesssim 10\, a_B$, see the insert. The correct short range asymptotic behaviour in $g(r)$ is reproduced only for $P\geq 8$. This result is expected as the corresponding high-temperature factorization, Eq.~(\ref{4thorder}), is optimized to be accurate up to the higher order contributions, $O(\epsilon^4), \, \epsilon=\beta/P\ll 1$, only for the internal energy~\cite{Sakkos2009}.}
\label{fig:gpair}
 \end{center}
\end{figure}

In order to keep the systematic errors due to the neglected high-order commutators smaller than the statistical QMC errors, i.e. the terms of the order $O(\epsilon^4)$ in Eq.~(\ref{4thorder}) which can be estimated from the Baker–Campbell–Hausdorff formula~\cite{Suzuki1985}, in Fig.~\ref{fig:gpair} we present the $P$-convergence test for main thermodynamic properties. 
The results for the internal energy components (see panels a,b) are well converged already for $P=2$ at temperature $\theta=1$ (the observed deviations are within the statistical error bars). In contrast, some $P$ dependence is still observable in the short-range correlation part of the radial distribution function (Fig.~\ref{fig:gpair}c). The similar analysis performed for the statistic structure factor $S(q)$ (SSF), being the central quantity for the estimation of the fourth frequency moment $C_4(q)$ (Eq.~(\ref{c4})), has confirmed that the factorization errors practically vanish for $P\geq 4$.

In summary, for the densities $r_s \geq 2$ and the temperatures  $1\leq \theta \leq 8$ ($\theta=T/T_F$), we end up with the optimal choice $P=8$. In particular, for the low density case ($r_s \geq 16$) analyzed in Sec.~\ref{DSFrs36}, the UEG degeneracy factor is relatively small ($n\lambda^3 \lesssim 0.1 $) and the average sign~(\ref{def_Sign}) only has a weak $P$-dependence. For $r_s=16$ and $N=34$ it varies within the range $0.52|_{P=16}\leq  \Avr{S(P)} \leq 0.63|_{P=16}$. 
In addition, we admit that the simulations with a larger value of $P$ are better suited for reconstruction of the dynamical properties as they deliver a more refined resolution of the intermediate scattering function $F(q,\tau)$ in the imaginary time, Eq.~(\ref{Fqw}). The latter is  used, in particular, for the accurate evaluation of the static density response function $\chi(q,0)$, see Eq.~(\ref{chi0}), and in the optimized reconstruction procedure for the higher-order power moments $C_6,C_8$  discussed in Sec.~\ref{SecReconstr}. We obtained well-converged results for $\chi(q,0)$ for $q\leq 6 q_F$ using both $P=8$ and $P=16$. The integral in Eq.~(\ref{chi0}) was performed using the spline interpolation between the values of  $F(q,\tau_p)$ resolved at the discrete argument values $\tau_p=p \epsilon$.

\begin{figure}[t]
\begin{center}
\hspace{-0.3cm}
\includegraphics[width=0.51\textwidth]{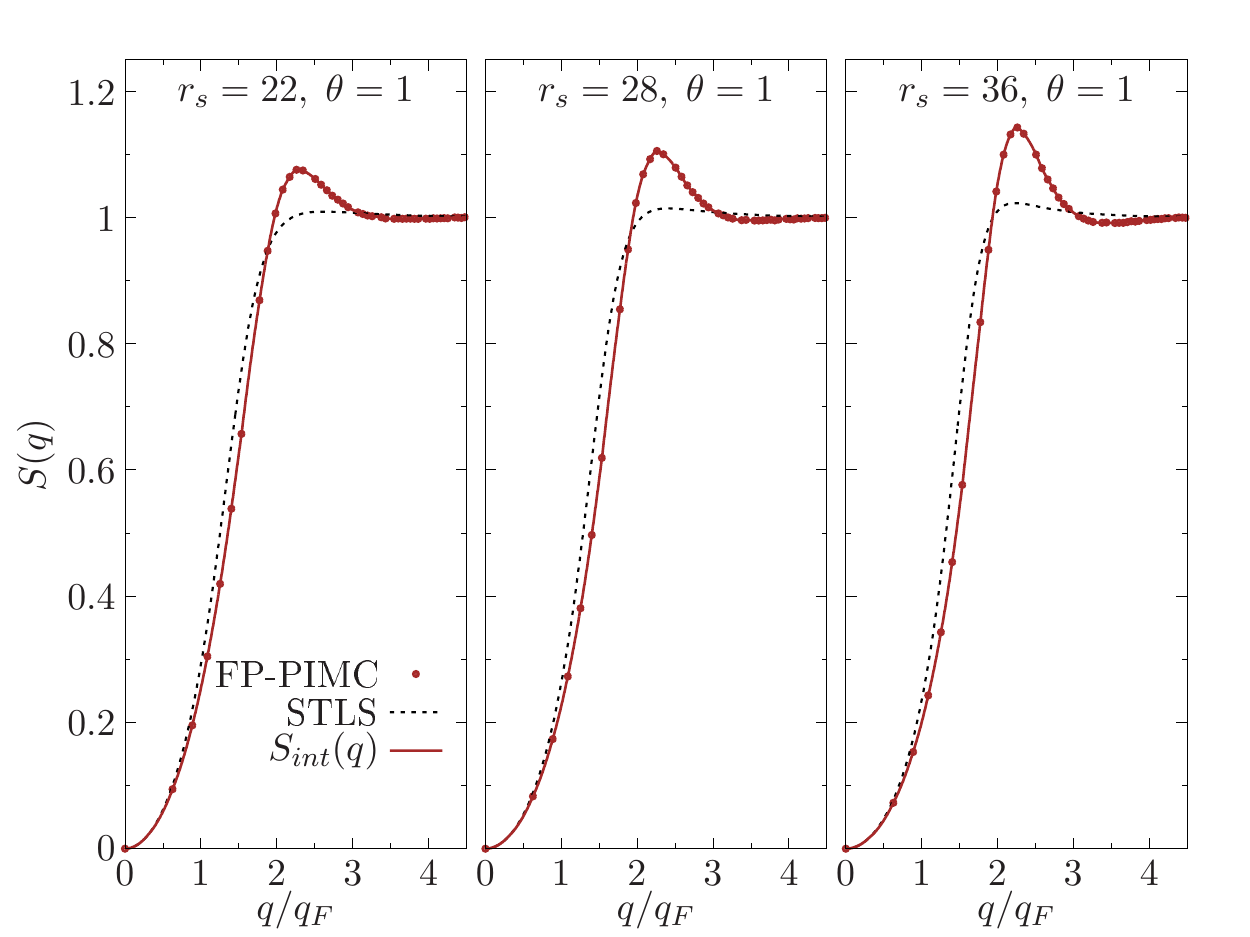}
\vspace{-0.7cm}
\caption{The static structure factor at $r_s=22,28,36$ and $\theta=1$. The solid line demonstrates the spline interpolation $S_{\text{int}}(q)$ over the FP-PIMC data (solid dots) available for $q\geq q_{\text{min}}\, (2\pi/L)$ and the STLS results~\cite{1986JPSJ} which provide the long-wavelength asymptotic behavior not accessible with the finite-size PIMC simulations ($q_{min}=0.627 q_F$ for $N=34$). For the detailed analysis of the validity of the STLS theoretical approach, see e.g. Ref.~\cite{filinov-etal.2021ctpp}.}
\label{fig:SkSTLS}
 \end{center}
\end{figure}

A similar spline interpolation procedure is required in the integral~(\ref{c4}) 
applied to the static structure factor $S(q_n)$ being defined only for the discrete set of momentum $q_n=2\pi n/L$ $(n=0,1,\ldots)$ with $N/L^3=(\frac{4}{3} \pi r_s^3)^{-1}$ due to a finite system size $N$ and the periodic boundary conditions (PBC). The values of $S(q)$ below the minimum wavenumber, $q_{\text{min}}=2 \pi/L$, have been complemented by the STLS theory~\cite{1986JPSJ} similar to the analysis presented in Ref.~\cite{filinov-etal.2021ctpp}. Some examples are presented in Fig.~\ref{fig:SkSTLS}.  

Finally, notice that we employed the standard periodic boundary conditions with the Ewald summation procedure~\cite{Fraser1996} to take into account the long-range nature of Coulomb interaction. While this allows to significantly reduce the finite-size effects in the static structural properties (or even make them negligible, see below), for the most important thermodynamic properties such as the internal energy and the free energy the corresponding scaling analysis should be conducted carefully~\cite{dornheim_physrep_18,filinov-etal.2021ctpp}. 

\subsection{Finite-size effects}

The predictions on the system dynamical response discussed in the next sections are based on the general expressions valid in the thermodynamic limit. However, the self-consistent method of moments introduced below employs as a crucial input the static properties evaluated in finite-size simulations. Therefore, their dependence on the system size $N$ has to be validated. This concerns, in the first place, the static density response function $\chi_N(q,0)$ and the static structure factor $S_N(q)$ which enter explicitly in the moments $C_0(q)$ and $C_4(q)$. 

\begin{figure}[t]
\begin{center}
\hspace{-0.3cm}
\includegraphics[width=0.51\textwidth]{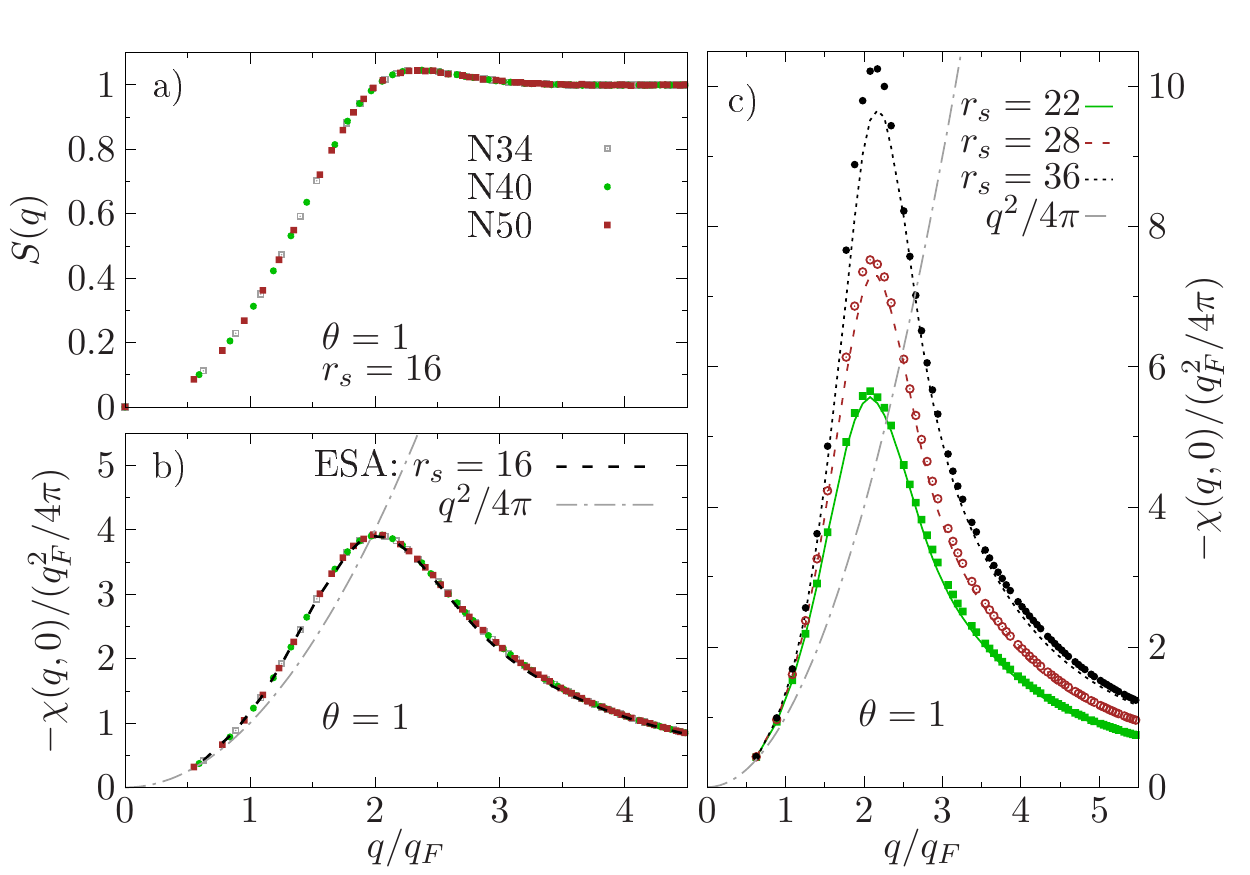}
\vspace{-0.7cm}
\caption{The finite-size dependence of the static structure factor $S(q)$ (panel (a)) 
and the static density density response function $\chi(q,0)$ (panel (b)) for $N=34,40,50$. 
The symbols corresponds to the allowed discrete values of the wavenumber, $q_n=2 \pi n/L_N \; (n=1,2,\ldots)$ due to the periodic boundary conditions. 
Simulations are performed for $r_s=16$, $\theta=1$ and $P=8$. Panel c: The lower density case: $22 \leq r_s\leq 36$. The $\chi(q,0)$ from the FP-PIMC ($N=34$) is compared to the effective static local field (ESA) result, $\chi_{\text{ESA}}(q,0)$, reconstructed via the LFC factor using the neural-net representation~\cite{PhysRevLett.125.235001}. For the reference, we include the expected long wavelength limiting asymptotic form, $\lim\limits_{q \rightarrow 0} \chi(q,0) =q^2/4 \pi$, due to a perfect screening condition in the UEG (dashed gray line).}
\label{fig:FSplot}
 \end{center}
\end{figure}

The results of simulations for both quantities are presented in Fig.~\ref{fig:FSplot}a,b for $N=34,40$ and $N=50$. Up to the statistical errors  we cannot resolve any finite-size effects present in our data. Our results are in agreement with the previous findings~\cite{DornheimFSE2021} for lower densities ($r_s < 10$). Next, for $r_s=16$ we validate our FP-PIMC data for $\chi(q,0)$ versus $\chi_{\text{ESA}}(q,0)$ evaluated via the static dielectric function in the RPA-type representation with the static local field correction taken from the neural-net representation~\cite{PhysRevLett.125.235001}.  The agreement is excellent up to $q\sim 3 q_F$. 

In Fig.~\ref{fig:FSplot}c we perform a similar comparison but for lower density case ($r_s\geq 22$). Since the neural net was trained only for $0.7 \leq r_s \leq  20$, we notice a very reasonable agreement at $r_s=22$, and observe some systematic deviations for larger $r_s$. The effective static approximation (ESA) results, in general, underestimate the amplitude of the main peak in $\chi(q,0)$, while both theoretical approaches converge to the same asymptotic limit for small $q$ given by the perfect screening sum rule in the UEG. To conclude, even though the ESA curves slightly deviate from the exact PIMC data, the observed deviations are not large, and the ESA approach is used further as a reference approximation where possible dynamical correlation effects in the density response are neglected. We note that this approach remains quite accurate at least for $r_s \leq 6$, see e.g. Ref.~\cite{PhysRevB.102.125150}.

\section{Extended self-consistent method of moments with dynamical correlations}\label{SCMM}

\subsection{Spectral density and frequency power moments.} 

From the mathematical point of view, the problem we solve in this work, is the \textit{truncated Hamburger problem of moments }
consisting in the reconstruction of a non-negative distribution density from its power moments~\cite{krein-book, akhiezer-book, shohat-book}.
This problem is solvable \cite{shohat-book} if and only if the Hankel matrices \cite{akhiezer-book} constructed from the moments are all non-negative. Certainly, if the distribution (spectral) density is an even function of frequency, the set of power moments and
the orthogonal polynomials which serve as the coefficients of the Nevanlinna
linear-fractional transformation \cite{nevanlinna-book, tkachenko-book} simplify significantly \cite{tkachenko-book}.

For this reason it is convenient to express all dynamical characteristics in terms 
of the loss function
\begin{equation}
\mathcal{L}(q,\omega)=-\frac{\operatorname{Im}\epsilon ^{-1}\left( q,\omega \right) }
{\pi \omega } \,  \label{lossf1}
\end{equation}%
which is non-negative by virtue of the fluctuation-dissipation theorem (FDT)
\begin{equation}
-\frac{\operatorname{Im}\epsilon ^{-1}\left( q,\omega \right) }{\pi \omega }=\frac{%
4\pi ne^{2}}{q^{2}}\frac{\left[1-\exp \left( -\beta \hbar \omega \right)\right]}{\hbar
\omega }S\left( q,\omega \right)  \label{fdt}
\end{equation}
and is an even function of frequency since $\operatorname{Im}\epsilon ^{-1}\left(q,\omega \right) $\ is an odd function of $\omega$.
The loss function frequency power moments
\begin{equation}
C_{\ell }\left( q\right) =\int\limits_{-\infty }^{\infty }\omega ^{\ell}\, \mathcal{L}(q,\omega)\, \db\omega \
,\;\; \ell=0,1,2,...,8, \label{lossf2}
\end{equation}%
and the characteristic frequencies determined by the sequential ratios of the power moments 
\begin{equation}
\omega_{j}\left( q\right) =\sqrt{C_{2j}/C_{2j-2}\left( q\right) }\ ,\quad
j=1,2,3,4\ ,  \label{oms}
\end{equation}%
will be the only construction blocks of the present approach. 

The odd-order moments vanish and the set of moments we consider simplifies 
into 
$\left\{ C_{0}\left(
q\right),0,C_{2},0,C_{4}\left( q\right) ,0,C_{6}\left( q\right)
,0,C_{8}\left( q\right) \right\} .$
Notice that the
frequency integral in Lindhard's formula for the polarizational stopping
power of a plasma is an incomplete second moment of the above loss function.

The static dielectric function, due to the Kramers-Kronig relations, is directly related to the zero-order moment, $C_{0}\left( q\right) =1-\epsilon ^{-1}\left(q,0\right)$; and the plasma frequency enters via the f-sum rule, $C_{2}=\omega_{p}^{2}$. 
The fourth moment by virtue of the detailed balance condition~\cite{ara-etal.2021pop} is effectively the third moment of the DSF and can be explicitly derived from the commutation relations~\cite{puff} and expressed as follows:  
\begin{eqnarray} 
&&C_4(q)=\frac{2 n}{\hbar}\, \Phi(q) \, \omega_0(q)\cdot \omega_3^2(q)=\omega_p^2\cdot \omega_3^2(q),\\
&& \omega_3^2(q)=\omega_0^2(q)+4 \omega_0(q)\cdot \epsilon_k/\hbar + \omega_p^2 [1-C_I(q)],\\
&& C_I(q)=\frac{1}{8 \pi^2 n} \int\limits_0^{\infty} \db k \, k^2 [1-S(k)] \cdot f(q,k) \, \label{c4} ,
\end{eqnarray}
where $\Phi(q)=4 \pi e^2/q^2$, $\omega_0(q)=\hbar q^2/2 m$. 
The factor
\begin{equation}
f\left( q,k\right) =\frac{5}{3}-\frac{k^{2}}{q^{2}}+\frac{\left(
k^{2}-q^{2}\right)^{2}}{4 k\, q^{3}}\ln \left\vert \frac{k+q}{k-q}\right\vert \,  \label{44}
\end{equation}
reflects the angular averaging in the momentum vector.

The fourth moment contains two main contributions: (i) the average kinetic energy per particle $\epsilon_k=\Avr{E_{\text{kin}}}/N$ \cite{iwamoto.1984prb} reduced in the case of a non-interacting system to the Fermi integral $I_{3/2}\left( \eta \right) $: $\epsilon_k^{\mathrm{ideal}} =3\theta ^{3/2}I_{3/2}\left( \eta \right)
/2\beta $, and (ii) the exchange-correlation contribution $C_{I}\left( q\right) $
with $S\left( q\right) $ provided, e.g., by the ab-initio QMC simulations~\cite{dornheim-etal.2017ccp}. In the present work, in order to access the region of small wavenumbers, $q \leq 0.6\, q_F$, dominated by a sharp plasmon resonance (see Fig.~\ref{fig:Ftau}) and higher values of the coupling parameter, $r_s \geq 16$, we performed an independent evaluation of the SSF with the fermionic propagator PIMC~\cite{filinov-etal.2021ctpp} and the system size such that  $64 \leq N\leq 140$. 

\subsection{The self-consistent solution of the five-moment problem. }

We start our analysis using a non-canonical solution of the {\it five}-moment Hamburger problem $\left\{ C_{0}\left( q\right),0,C_{2},0,C_{4}\left( q\right) \right\} $. 
The Nevanlinna theorem~\cite{nevanlinna-book, shohat-book, krein-book, akhiezer-book, tkachenko-book} establishes the following one-to-one linear-fractional transformation between the inverse dielectric function 
\begin{equation}
\epsilon^{-1}\left( q,\omega ;Q_{2}\right)=1+\frac{\omega_{p}^{2}\left(
\omega +Q_{2} \right)}{%
\omega \left( \omega ^{2}-\omega_{2}^{2}\left( q\right) \right)
+Q_{2}\left( \omega
^{2}-\omega_{1}^{2}\left( q\right) \right) }\ ,  \label{idf}
\end{equation}%
and a non-phenomenological Nevanlinna (response) function $Q_{2}=Q_{2}\left( q,\omega\right)$ such that $ \lim_{z\rightarrow \infty } Q_{2}\left( q,z\right) /z=0$ ($\text{Im} z>0$), see Ref.~\cite{krein-book}.
These solutions have been extensively tested against the molecular-dynamics simulations of classical one-component Coulomb and Yukawa systems~\cite{arkhipov-etal.2017prl, arkhipov-etal.2020pre} with the quantitative agreement achieved even within the static approximation for $Q_{2}\left( q,z\right)$, i.e. when
\begin{eqnarray}
Q_{2}\left( q,z\right)=\lim_{z \to 0^{+}}  Q_{2}\left( q,z\right) =ih_{2}\left( q;\omega_{1},\omega_{2}\right) \ .
\label{sa}
\end{eqnarray}
Since in the DSF of the above classical systems a broad extremum was observed at the zero frequency, the third derivative test for even functions~\cite{arkhipov-etal.2017prl, arkhipov-etal.2020pre} was applied to obtain the static Nevanlinna parameter 
$h_{2}\left( q;\omega_{1},\omega_{2}\right) =\omega_{2}^{2}\left( q\right)
/\left( \sqrt{2}\omega_{1}\left( q\right) \right) \ .$ 
Notice that the Nevanlinna function is directly related to the dynamic local field correction used to extend the random-phase approximation (RPA)~\cite{PhysRevE.81.026402}, see also~\cite{arkhipov-etal.2020pre}.

\subsection{The dynamic (five-moment) Nevanlinna parameter function. }

This approach being very accurate for classical systems proves to be insufficient for Fermi fluids, where the trimodal structure of the spectrum~\cite{takada.2016prb} and a significant shift with respect to the RPA plasmon~\cite{dornheim.2018prl} have been recently predicted. 

It has long been known~\cite{GNPSz} that the tri-modal spectrum (the zero-frequency mode plus two "shifted" modes) presumably should be attributed to the dynamical multi-pair effects in electron fluids, and can be described only when the local field becomes a complex dynamic function of the energy transfer $\hbar \omega$, in other words, if we abandon the static approximation~(\ref{sa}) for the five-moment Nevanlinna function and specify the high-frequency asymptotic behavior of the inverse dielectric function (IDF), which is a genuine response (Nevanlinna) function~\cite{krein-book}. To this end we equalized the five-moment expression for the IDF to the one stemming from the nine-moment solution
of the Hamburger problem taking into consideration the sixth and the eighth frequency moments, $C_{6(8)}$, or the frequencies $\omega_{3(4)}$ defined in Eq.~(\ref{oms}). Thus, we expressed the dynamic {\it five}-moment Nevanlinna function in terms of the {\it nine}-moment one and, using the same physical considerations~\cite{arkhipov-etal.2017prl,arkhipov-etal.2020pre} employed for the latter the static approximation similar to~(\ref{sa}). This construction is presented in detail in Appendix. Hence, the dynamic response problem was reduced to the study of only two new static characteristics which are the unknown frequencies $\omega_{3(4)}\left( q\right)$. Notice that the static nine-moment Nevanlinna parameter $h_{4}\left( q;\tilde\omega\right)$ with $\tilde\omega=\{\omega_1,\omega_{2},\omega_{3},\omega_{4}\}$ is determined by the frequencies $\omega_{3(4)}\left( q\right)$, since the frequencies 
$\omega_{1}\left( q\right) \ $and $\omega_{2}\left(q\right)$ are uniquely defined by the SSF and by the static density response function which is directly accessible from the intermediate scattering function:
\begin{eqnarray}
\chi(q,0)= - n\int\limits_0^{\beta}\db \tau\, F\left (q,\tau\right)=\frac{q^2 (\epsilon^{-1}(q,0)-1)}{4 \pi e^2}.\label{chi0}
\end{eqnarray}
We understand that the frequencies $\omega_{3(4)}\left( q\right)$, formally introduced above, are determined by the three- and four-particle static correlation functions. The \textit{ab initio} QMC data for them can be achieved though precise expressions for the higher-order moments in terms of these correlation functions not yet available. The precision of the latter seems to be a problem and, as we show, to achieve quantitative agreement with the simulation data, we need to possess highly precise values of the sixth and the eighth moments. This is why we determine their values by means of the Shannon information entropy maximization (EM)
procedure~\cite{shannon.1948,khinchin.53umn,zubarev-book,jaynes.1957pr}, see also \cite{tkachenko-book}, 
or using the intermediate scattering function, see below.

\subsection{The Shannon entropy maximization technique. } 

We introduce the two-parameter Shannon entropy functional defined  by the loss function spectral density:
\begin{eqnarray}
&&\mathcal{E}\left(q;\tilde\omega\right) 
=-\int\limits_{-\infty}^{\infty}\mathcal{L}\left(q,\omega;\tilde\omega\right) \ln \left[ \mathcal{L}\left(q,\omega;\tilde\omega\right) \right]
\db\omega \ .
\label{sh0}
\end{eqnarray}%
and resolve the corresponding maximization problem with respect to
$\omega_{3(4)}(q)$, with $\omega_{1(2)}(q)$ fixed by the known sum-rules. 
To solve the extremum conditions for two unknown frequencies 
\begin{eqnarray}
\int\limits_{-\infty }^{\infty } &&\left\{ \frac{\partial \mathcal{L}\left(
q,\omega;\tilde\omega\right) }{\partial \omega_{3(4)}}\ln \left[
e \mathcal{L}\left(q,\omega;\tilde\omega\right) \right] \right\}
\db\omega=0 \ ,
\label{sh1}
\end{eqnarray}%
we employ the Newton-Raphson method. As the starting points in the gradient descent method the corresponding Fermi-Dirac distribution moments
\begin{equation*}
\omega _{30}\left( q\right) =\sqrt{\frac{I_{5/2}(\eta )}{I_{3/2}(\eta )}}%
\Omega _{s}\left( q;\eta \right) \ ,\quad \omega _{40}\left( q\right) =\sqrt{%
\frac{I_{7/2}(\eta )}{I_{5/2}(\eta )}}\Omega _{s}\left( q;\eta \right) \ ,
\end{equation*}%

have been chosen with
\begin{equation*}
\Omega _{s}\left( q;\eta \right) =\sqrt{\frac{I_{1/2}(\eta )}{I_{3/2}(\eta )}%
}\omega _{2}\left( q\right) \ .
\end{equation*}%
The Hessian of the entropy~(\ref{sh0}) was studied to warrant the satisfaction of the maximization condition.

\subsection{The eigenmodes and the dynamic structure factor: Comparison to
the local-field-based approach.}\label{Sec2C} 

Within our approach the properties of the eigenmodes can be directly studied via the solution of the dispersion equation, 
i.e. as the poles of the inverse dielectric function~(\ref{idf}). The corresponding algebraic equation is of the fifth-order:
\begin{eqnarray}
z\left( z^{2}-\omega_{2}^{2}\left( q\right) \right) +Q_{2}\left( q,z\right)
\left( z^{2}-\omega_{1}^{2}\left( q\right) \right) =0 \ .  \label{deQ2}
\end{eqnarray}
Hence, we obtain five complex frequencies
\begin{eqnarray}
&&z_{0}\left( q\right) =-i \Delta \Omega_0\left( q\right),\, \Omega_0=\text{Re}(z_0)=0,\label{diss0}\\
&&z_{\pm 1 (\pm 2)}\left( q\right) =\pm \Omega_{1(2)}(q)
-i\Delta \Omega_{1(2)}\left( q\right),\label{diss12} 
\end{eqnarray}%
which correspond to three possible eigenmodes: the diffusion (or Rayleigh) mode $\Omega_0(q)$ and two shifted modes $\Omega_{1(2)}(q)$. The intrinsically negative imaginary parts of the solutions are defined by the decrements of the corresponding modes, $\Delta \Omega_0\left( q\right) $ and $\Delta \Omega_{1(2)}\left( q\right) $.

\begin{figure}[t]
\begin{center}
\hspace{-0.3cm}
\includegraphics[width=0.51\textwidth]{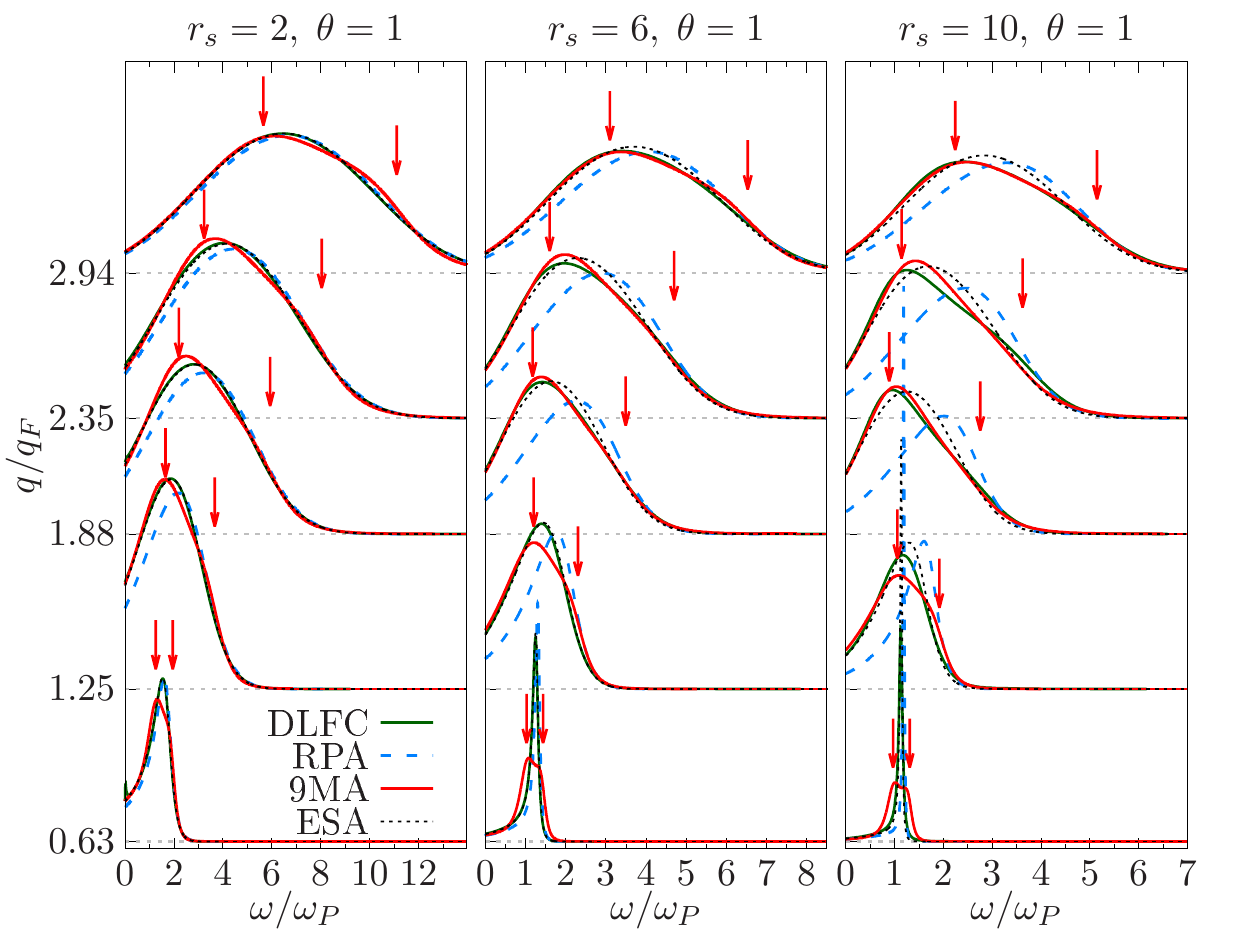}
\vspace{-0.3cm}
\caption{The dynamic structure factor $S(q,\omega)$ at three densities $\{r_s=2,6,10\}$ and temperature $\theta=1$. The frequency is normalized to the plasma frequency, $\omega/\omega_p$. The DSF plots are shifted by the value of the dimensionless wavenumber $q/q_F$ (horizontal dotted lines) with $q_F$ being the Fermi wavenumber, $\hbar q_F=\sqrt{2 m E_F}$. Compared are the results of the random-phase approximation (RPA), the {\it effective static approximation} (ESA)~\cite{PhysRevLett.125.235001}, the dynamic local field (DLFC)~\cite{dornheim.2018prl}, and the present self-consistent method of moments in the {\it nine}-moment (9MA) approximation. Qualitative discrepancies observed for $q\approx 0.63 q_F$ are due the Shannon entropy maximization which tends to smooth sharp energy resonances and is much better suited for the description of a broad multi-excitation continuum. 
Red vertical arrows indicate the frequencies $\Omega_{1(2)}(q)$ for the set of wavenumbers, $q/q_F=\{0.6269,1.2538,1.8808,2.3457,2.9405\}$, specified by the periodic boundary conditions for $N=34$, i.e. $q=\sqrt{q_1^2+q_2^2+q_3^2}$ with $q_{i}=2\pi n_i/ L$ ($n_i=1,2,\ldots$).}
\label{fig:skdynb1}
 \end{center}
\end{figure}

We applied the present self-consistent method of moments in the {\it nine}-moment approximation (9MA) to reconstruct $S(q,\omega)$ for different sets of parameters $\{r_s,\theta\}$.
The performance of our approach in the WDM regime is demonstrated in Fig.~\ref{fig:skdynb1} where it is compared to the DLFC results~\cite{dornheim.2018prl,groth.2019prb}. 
Both methods are in a good quantitative agreement both for weak ($r_s=2$) and moderate ($r_s=6,10$) coupling, as they directly include the exchange-correlation contribution~(\ref{c4}). The positions of the maxima and their broadening due to the damping are reproduced very accurately. The damping effects are intrinsically present in the 9MA solution due to the dynamical nature of the {\it five}-moment Nevanlinna function.

\begin{figure}[t]
\begin{center}
\hspace{-0.3cm}
\includegraphics[width=0.51\textwidth]{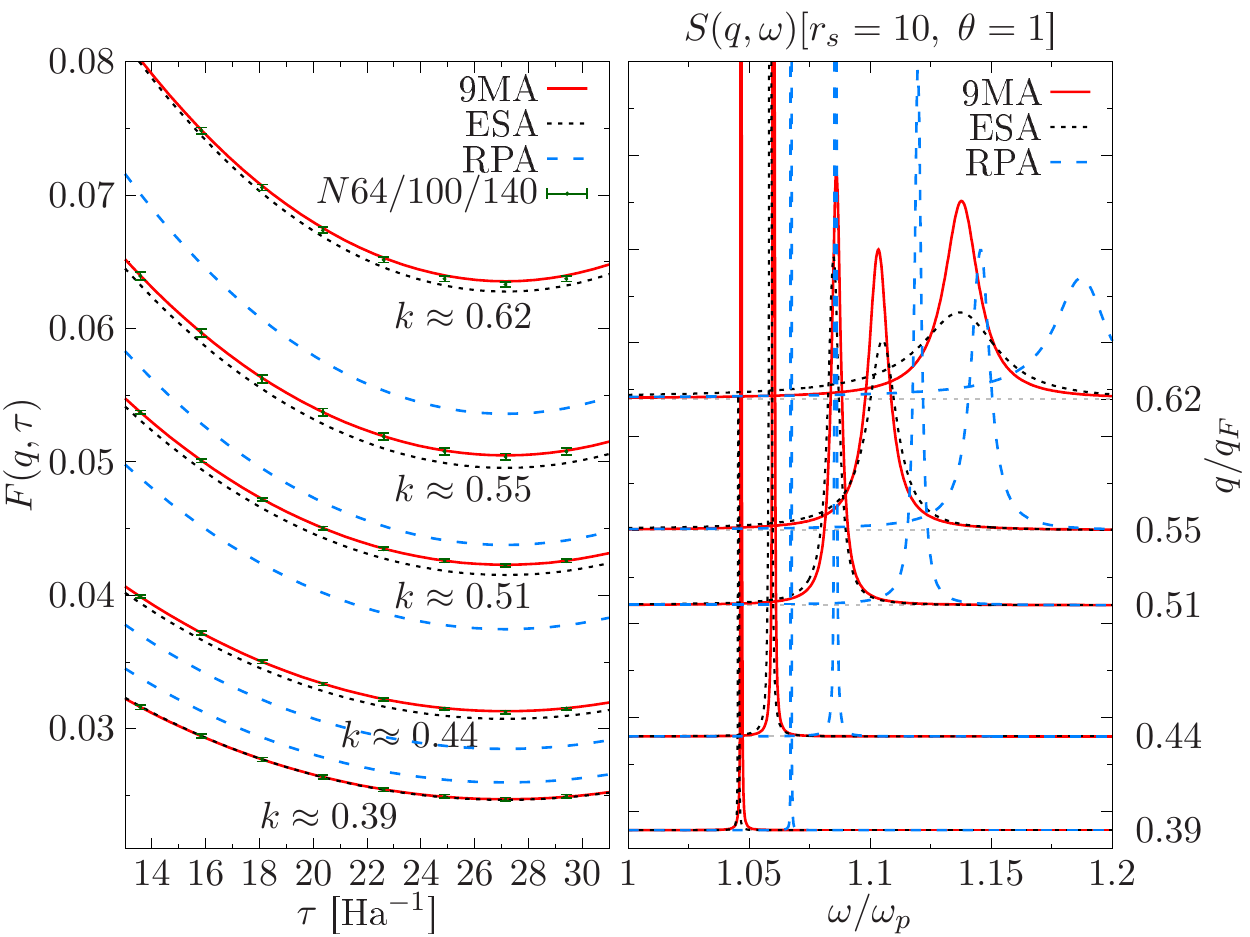}
\vspace{-0.5cm}
\caption{{\it Left}: The intermediate scattering function, $F(q,\tau)$, for $r_s=10$ ($\theta=1)$ and the wavenumbers $k=q/q_F=\{0.3911,0.4376,0.5077,0.5532,0.6188\}$, corresponding to $q=2\pi n_i/L_N \, (n_i=1,2)$ from the PIMC simulations~\cite{filinov-etal.2021ctpp} for different system sizes, $N=64,100,140$ (symbols with the error bars). The function $F(q,\tau)$ is symmetric with respect to $\tau_0$, ($\tau_0=\beta/2=27.15$ Ha$^{-1}$) and its initial value ($\tau=0$) is defined by the SSF, $F(q,0)=S(q)$.
{\it Right}: The reconstructed dynamic structure factor, $S(q,\omega)$, within several approximations: the RPA, the ESA~\cite{PhysRevLett.125.235001}, and the present  9MA self-consistent method of moments. The DSF plots are shifted by the value of $q/q_F$ (horizontal dotted lines). The agreement of $F(q,\tau)$ is evaluated via the Laplace transform of $S(q,\omega)$ (from RPA, ESA, 9MA), see Eq.~(\ref{Fqw}), with the ab initio PIMC data provided on the left panel. The DLFC results~\cite{dornheim.2018prl,groth.2019prb} (see Fig.~\ref{fig:skdynb1}) are not available for $q< 0.63$ ($N > 34$).} 
\label{fig:Ftau}
 \end{center}
\end{figure}

The only case when the DLFC results become qualitatively different from the 9MA ones is $q \approx 0.63 q_F$ ($r_s=6,10$), where only a single sharp plasmon resonance quite accurately reproduced within the RPA and the ESA is present.
At these conditions, the Shannon EM provides a class of solutions which are too smooth, and, hence, any sharp resonance features, if present in the spectrum, are artificially broadened, though the spectral density still satisfies all imposed constrains including the five lower-order moments
$\left\{ C_{0}\left( q\right),0,C_{2},0,C_{4}\left( q\right) \right\} $ which are known {\em exactly} from the Monte-Carlo data. To avoid such artefacts induced by the unknown
higher moments $C_{6(8)}$, we re-evaluated the DSF with the frequencies $\omega_{3(4)}$ used as the fitting parameters to reproduce the decay of $F(q,\tau)$, see Eq~(\ref{Fqw}), obtained within the fermionic PIMC~\cite{filinov-etal.2021ctpp}. 
Thus we found a much better agreement with the DLFC data at $q\approx 0.63 q_F$. We applied this idea for smaller wavenumbers beyond the DLFC-generated data. These new results are presented in Fig.~\ref{fig:Ftau} and clearly demonstrate the applicability of our analytical expression for the inverse dielectric function~(\ref{idf}) even in the case of a sharp resonance, see the DSF for $q\approx 0.39 q_F$ in Fig.~\ref{fig:Ftau}. The accuracy of the reconstructed $S(q,\omega)$ is justified by the agreement with $F(q,\tau)$. Among other approximations (RPA, ESA) only the 9MA agrees with the intermediate scattering function within the statistical error bars (see the left panel in Fig.~\ref{fig:Ftau}). The details of this approach will be discussed in Sec.~\ref{SecReconstr}.           

To summarize, a simple combination of the fitting procedure with only two parameters in the case of sharp energy resonances (at lower $q$) and the {\em dynamical} approximation for the Nevanlinna parameter function satisfying the Shannon EM principle when the damping effects prevail allowed us to reproduce the UEG DSF in a broad range of variation of the momentum and at different densities with a high accuracy. The transition between both regimes can be physically justified  by a drastic variation of the decrement of the plasmon mode once it enters into the pair-continuum region~\cite{PN-book}. 
In particular, for $r_s=2$, the lower dispersion curve $\Omega_1(q)$ obtained from the solution~(\ref{diss12}) at $q \approx 0.63 q_F$ (see the first panel in Fig.~\ref{fig:modes}) already lies at the edge of the pair continuum, and the present approach provides a very accurate description. Other theories (RPA, ESA) demonstrate here a similar accuracy. The ESA theory is based on the static LFC and for the weak coupling ($r_s=2$) it leads to a nearly perfect agreement with the DLFC data for all wavenumbers. In contrast, the RPA prediction becomes unreliable in a finite interval,  $1.2 \lesssim q/q_F \lesssim 2.9$, when the account of the static pair correlations becomes necessary via the $G(q,0)$-factor, as demonstrated in the STLS theory~\cite{STLS}. The validity of the RPA solution is restored once the ESA LFC approaches unity for large wavenumbers. 

Similar trends are observed for $r_s=6(10)$. Omitting the case of a sharp plasmon resonance, the best agreement with the DLFC for $q/q_F\geq 1.88$ is provided by the 9MA reconstruction. The asymmetric form of the DSF and a noticeable {\it redshift} of its maximum with respect to the RPA/ESA results indicated by a vertical arrow (see the second (third) panel in Fig.~\ref{fig:skdynb1}) are reproduced quite well. 

On the contrary, we observe systematic deviations (more pronounced for $r_s=10$) between the ESA and the DLFC models. The onset for this discrepancy matches the characteristic wavenumber $q_c$ when the dispersion curve, $\Omega_1(q)$, in the second (third) panel in Fig.~\ref{fig:modes} enters the pair excitation region: $q_c\sim 0.62 q_F$ ($r_s=2$), $q_c\sim 0.9 q_F$ ($r_s=6$), and  $q_c\sim 1.0 q_F$ ( $r_s=10$). For $q > q_c$ we observe that the reconstruction with the dynamical Nevanlinna parameter function starts to demonstrate a remarkable agreement with the DLFC and the PIMC data for $F(q,\tau)$. This testifies the importance of the dynamical correlations and the need for the dynamical local field theory in this regime substituting the static LFC approximation used in the ESA. 

These observations validate the physical consistency of the applied Shannon EM technique at high/moderate densities. Once the plasmon mode is strongly damped and broadened, one observes that the spectral density is mainly formed by the contribution of different combinations of quasi-particle excitations – the microstates in the sense of the statistical ensemble. The most probable (degenerate) solution in this case should correspond to the entropy maximum. This permits to determine the unknown frequencies $\omega_{3(4)}$ by means of the Shannon EM extrema conditions in a unique way.

Furthermore, the frequencies of the eigenmodes, $\Omega_{1(2)}(q)$, found as the poles of the inverse dielectric function, are also compared in Fig.~\ref{fig:skdynb1} to the full DSF results. We observe a quite good agreement between the low-frequency mode $\Omega_{1}(q)$ and the maximum of the spectral density (excluding $q\approx 0.63 q_F$). 
The second solution $\Omega_{2}(q)$ is shifted to higher frequencies, and in our interpretation (see below) it is responsible for the observed asymmetrical shape of the DSF. This effect becomes more pronounced at low  densities ($r_s \gtrsim 6$), when
we can observe even a second local maximum predicted independently (Fig.~\ref{fig:skdynb1}) both by the DLFC model at $q/q_F \approx 2.35,2.94$ ($r_s=10$), and within the 9MA theory at $q \approx 2.94 q_F$ ($r_s=2,6,10$) and $q \approx 1.25 q_F$ ($r_s=6,10$). Both approaches indicate the presence of two modes, distinguishable at low and high frequencies, which, however, are difficult to resolve if only a full DSF is available.

Recently, the dispersion relation, $\epsilon(q,z)=0$, has been analyzed on the complex frequency plane within the ESA approximation~\cite{hamann-etal.2020CPP} based on ab initio QMC data for the static LFC. Only a single solution (a plasmon) was found and only outside the $q-\omega$ region corresponding to the pair continuum (Fig.~\ref{fig:modes}). In contrast, within the present approach the possibility for a three-mode solution (including a diffusive mode) and the mode-mixing effects are incorporated in the analytical representation of the Nevanlinna parameter function and the inverse dielectric function.  
Our solution for the dispersion relations~(\ref{diss12}) of two shifted characteristic modes $\Omega_{1(2)}(q)$ is demonstrated in Fig.~\ref{fig:modes}. 
In the WDM regime ($r_s=2$) the $\Omega_1(q)$ mode lies close to the center of the pair continuum. A similar behaviour is observed for the ESA/RPA but the deviations increase with $r_s$. For $r_s=6(10)$ we clearly observe the negative dispersion and a local roton-like feature in the range $1.5 \lesssim q/q_F \lesssim 2.5$. The effect is more pronounced compared to the ESA predictions and is in a good agreement with the DLFC~\cite{dornheim.2018prl} results.
Around $q\sim 1.9 q_F$ the strongly damped $\Omega_1(q)$ mode is responsible for the low-frequency DSF maximum, while the upper branch $\Omega_2(q)$ generates a broad shoulder at higher frequencies. Moreover, around $q \sim q_c$ this shoulder is centered close to the RPA dispersion. A similar behavior is captured quite well also by the DLFC for $q/q_F \approx 2.35,2.94$ ($r_s=10$) visible in Fig.~\ref{fig:skdynb1}. Our analysis performed for small wavenumbers ($q< 0.63 q_F$, see Fig.~\ref{fig:Ftau}) has proved that the upper branch $\Omega_{2}(q)$ in the long-wavelength limit coincides with the plasmon mode.

Furthermore, the lower branch, $\Omega_{1}(q)$, was found to exist only in the $q-\omega$ region spanned by the pair-continuum and has a negligible spectral weight in the full DSF (see Fig.~\ref{fig:Ftau}) when the upper mode, $\Omega_{2}(q)$, forms a sharp plasmon resonance. However, with the increase of the plasmon damping with the wavenumber $q$, the $\Omega_1$-mode contribution is systematically enhanced. In particular, the Shannon EM applied at $q\approx 0.63 q_F$ predicts a nearly equal spectral weight of both modes (see the DSF in the first row of Fig.~\ref{fig:skdynb1}). The intermediate scattering function $F(q,\tau)$ reconstructed from the ESA, the DLFC and the 9MA coincides with the PIMC data within the statistical error bars, and, therefore, cannot be used as a sufficient criteria to select a unique physical solution. For larger $q$ ($q>0.63 q_F$) all theoretical approaches, except RPA, predict the DSF maximum being close to $\Omega_1(q)$.

\subsection{Intermediate analysis of the UEG eigenmodes}\label{interAnal}

\begin{figure}[t]
\hspace{-0.9cm}
\includegraphics[width=0.515\textwidth]{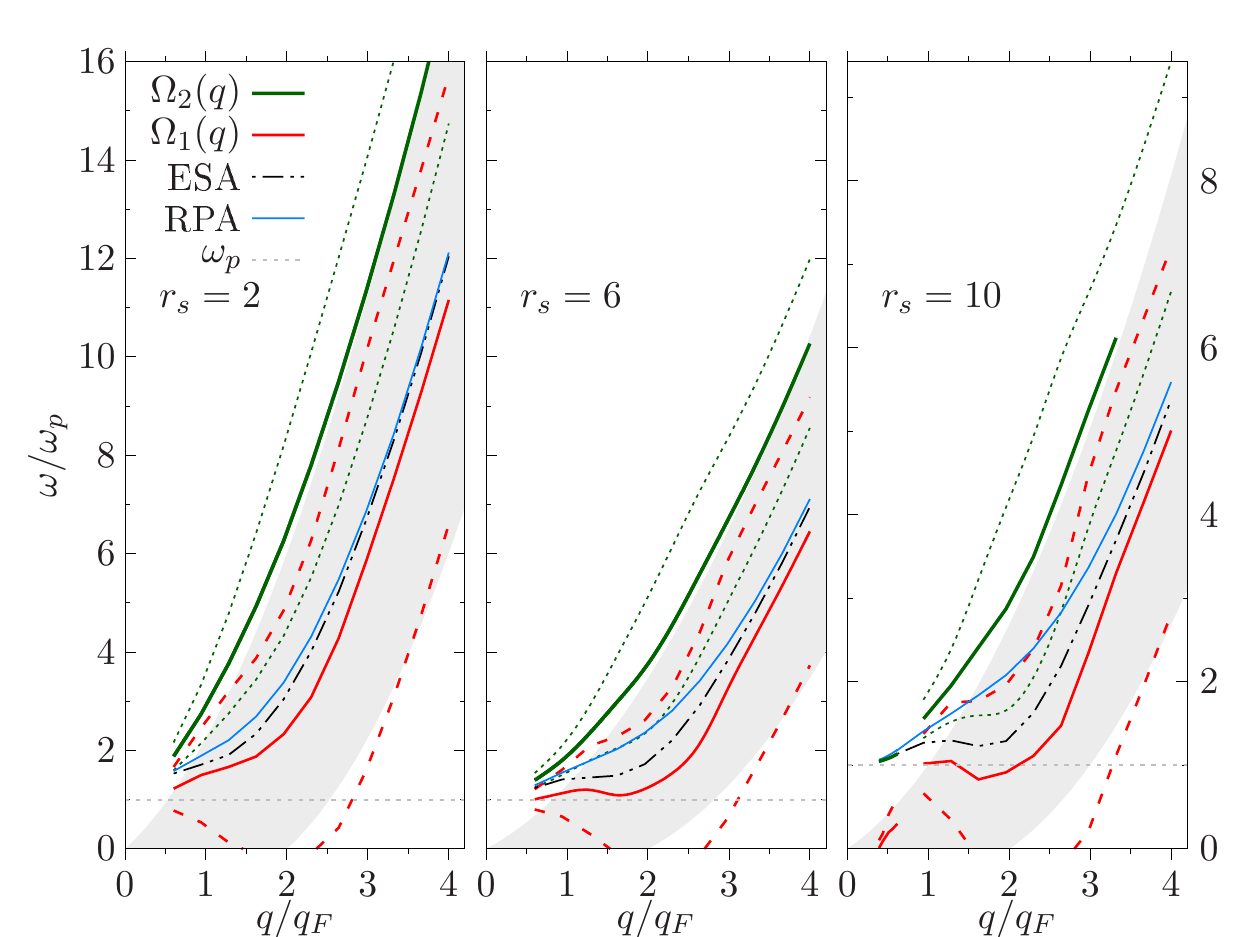}
\vspace{-0.3cm}
\caption{The wavenumber dependence of the solutions of the explicit dispersion equation, $\epsilon \left( q,z\right)=0$. Simulation parameters: $\theta=1$ and $r_s=2,6$ and $10$ [shown on the right most panel with a different scaling]. Two solid curves denote the modes: $\Omega_{1}(q)$ (red) and $\Omega_{2}(q)$ (green). The corresponding dashed lines $\Omega_{1(2)}(q)-\Delta \Omega_{1(2)}(q)$ and $\Omega_{1(2)}(q)+\Delta \Omega_{1(2)}(q)$, where $\Delta \Omega_{1(2)}(q)$ are the decrements, delimit the linewidths. For $r_s=10$ is shown, in addition, the dispersion for $q/q_F\in [0.39,0.62]$ as resolved from Fig.~\ref{fig:Ftau}, where $\Omega_2(q)$ nearly coincides with the ESA prediction when the lower branch $\Omega_1(q)$ has a negligible spectral weight. Two dispersion equation solutions exhibit significant broadening when they approach the pair excitation continuum (the shaded area): $\hbar \omega \in  [\epsilon_{q+q_F}-\epsilon_{q_F},\epsilon_{q-q_F}-\epsilon_{q_F}]$ with $\epsilon_q=\hbar^2  q^2/2m$. Notice that the second solution $\Omega_{2}(q)$ lies near the edge of the pair excitation continuum and has a slightly reduced decrement compared to the lower one, $\Delta \Omega_{2}< \Delta \Omega_{1}$. The lower solution $\Omega_{1}$ always stays within the pair continuum. In the range $q/q_F\in [0.62,0.94]$ a unique dispersion cannot be resolved as both of the solutions, i.e the $F(q,\tau)$ fitting and the Shannon EM, reproduce the intermediate scattering function within the QMC error bars.
 The positions of the DSF maxima deduced from the RPA/ESA (solid blue/dashed black lines) are included for comparison.
}
\label{fig:modes}
\end{figure}

The above analysis of characteristic collective modes in electronic fluids or the UEG at moderate densities ($\theta=1$, $r_s=2,6,10$) within the {\it nine}-moment approximation complemented with the ab initio QMC data can be summarized as follows. We clearly observe how the position of the DSF peak undergoes a transition from the $\Omega_2(q)$ plasmon for $q < q_c$ (outside the pair continuum region) to the strongly damped low-frequency branch $\Omega_1(q)$ when the plasmon can decay into pair excitations. The main effect introduced by the exchange-correlation contribution $C_I(q)$ in the $C_4$-moment~(\ref{c4}) is the formation of a roton-like feature missing in the RPA theory completely.

For $q > q_c$ our dispersion equation~(\ref{deQ2}) predicts the presence of an additional second mode $\Omega_2(q)$, evolved from the plasmon for $q < q_c$, but with a significantly enhanced decrement $\Delta \Omega_2$. 
We believe that due to a strong damping it is not of the collective nature and can be viewed as a local enhancement of the spectral density around the Fermi energy. 
In addition, the upper edge of the pair continuum in Fig.~\ref{fig:modes} (shown at $\theta=0$) will be broadened at the simulated temperature $\theta=1$. 
The presence of both characteristic modes is practically indistinguishable in the full DSF as they strongly overlap due to a rapid increase of the corresponding decrements $\Delta \Omega_{1(2)}(q)$ whose role is represented in Fig.~\ref{fig:modes}, see the red (green) dashed  curves. 

As it will be demonstrated below, the role of the second solution $\Omega_2(q)$ interpreted here as a local maximum in the multi-excitation continuum can change at different thermodynamic conditions. In particular, at much lower densities ($r_s\geq 16$) and temperatures, it can acquire a collective character being a combination of a several quasiparticle excitations with a significantly long lifetime. These new physical predictions are discussed in detail in Sec.~\ref{Corr_effects}.

As to the possible physical interpretation of the $\Omega_{1}(q)$ mode when it is strongly overdamped ($\Delta \Omega_1 \sim \Omega_1$), its true physical origin has not yet been sufficiently clarified. The red-shift in the DSF maximum around $q\sim 2 q_F$ at metallic densities ($r_s\sim 4$) is a real physical effect and has been observed experimentally in alkali metals~\cite{Felde1989} and aluminium~\cite{takada.2002prl}. Takada~\cite{takada.2005,takada.2016prb} in his theoretical analysis attributed the roton-like feature to the excitonic mode dominant in the spectrum around $q \sim 2 q_F$.
The predicted excitonic mode has a two-particle character (an electron-hole excitation) and, therefore, it is mostly pronounced in the wavenumber segment spanned by the pair continuum. The idea of existence of such a mode in UEG has been discussed in a number of papers~\cite{weisskopf,himpsel,FLL.2022}. According to this concept, in order to conserve charge and angular momentum, an exchange electron is added to the exchange hole, forming a neutral exchange exciton. Consequently, the pair correlation defining the exchange hole is generalized to a three-fermion correlation. Certainly, this effect does not exist in classical systems.
Recently, Dornheim {\it et. al.}~\cite{align_model2022} provided an alternative microscopic explanation of a roton feature in terms of an electronic pair alignment model. It was qualitatively demonstrated that the maximum of the RPA-based spectral density should get a shift to lower frequencies due to the exchange-correlation correction in the potential energy part of the quasiparticle excitation, $\omega(q)= \omega_{\text{RPA}}(q)-\alpha \Delta W_{XC}(q)$. Still, the presented model was not capable of predicting the explicit form of the DSF and how it could be modified due to the quasiparticle interaction and damping effects. 

In summary, both theory trends underline the leading role of short range correlations either in electron-hole pairs (excitons) or electron pairs (two-particle alignment).
Leaving the physical interpretation of the "rotonization" of the spectrum as a collateral question, in the following analysis we will concentrate on a physically reliable and accurate reconstruction of the full DSF, and report a new evidence on even more pronounced roton-feature observed in the low-density UEG in Sec.~\ref{DSFrs36},~\ref{DSFDis}.

\section{Correlation effects in the dynamical response}\label{Corr_effects}

As it is discussed earlier, with the introduction of the dynamical Nevanlinna parameter function we are able (i) to reproduce the dynamical correlations in the DSF on the same level of accuracy as the dynamical local field~\cite{dornheim.2018prl,groth.2019prb, PhysRevB.102.125150}, and (ii) to observe a high-frequency mode, which generates a high-frequency shoulder, most pronounced at lower densities, $r_s=10$. Moreover, the direct solution of the dispersion equation, $\epsilon(z,q)=0$, permits to predict that the characteristic frequency of this mode lies slightly above the double plasmon frequency, i.e. $\Omega_2(q) \geq 2 \omega_p$, see Fig.~\ref{fig:skdynb1}. However, a clear observation of this mode in the full DSF is difficult due to strong damping effects in the density regime presented in Fig.~\ref{fig:modes}: the linewidths of two modes, $\Delta \Omega_{1(2)}(q)$, overlap strongly.        

Motivated by these observations, we extend our 9MA approach to lower densities, i.e. consider the  UEG dynamical characteristics at $\{r_s=16,22,28,36\}$, where the Coulomb correlations dominate. The use of the five-moment dynamical Nevanlinna function allows for  an ab initio reconstruction of the DSF of electron fluids including dynamical correlations at these conditions for the first time.

Since the existing results employing the dynamic local field are limited to intermediate coupling, $r_s\leq 10$, for a valuable comparison we use the simulation data based on the effective static local-field correction (ESA) reconstructed at the same thermodynamic conditions $\{r_s,\theta\}$ using the neural network representation~\cite{PhysRevLett.125.235001}. The corresponding static local-field factor, $G(q)$, proceeding from the ab initio QMC data contains full information of the static correlations in the system.     

For these new studies we have performed the fermionic PIMC simulations~\cite{filinov-etal.2021ctpp} with the temperature varied in the range, $1\leq \theta \leq 8$. Notice that due to the definitions, $\theta=T/T_F$ and $T_F=(\hbar^2/2m) (3 \pi^2 n)^{2/3}\sim r_s^{-2}$, by increasing the coupling parameter from $r_s \sim 10$ to $r_s\sim 36$ we achieve to diminish the physical temperature by a factor of $13$. Hence, the DSF results presented below at $\theta=1$ demonstrate a low-temperature counterpart of the excitation spectrum in Fig.~\ref{fig:skdynb1} with significantly suppressed thermal effects. The physical temperature becomes comparable with that in Fig.~\ref{fig:skdynb1} for $\theta\sim 2.5$ $(r_s=16)$, $\theta\sim 5$ $(r_s=22)$ and $\theta\sim 8$ $(r_s=28)$. The plasmon frequency is reduced with the density as well, however, the corresponding reduction is weaker since $\omega_p\sim r_s^{-3/2}$. Hence, the thermal contribution to the damping will be scaled as, $k_B T/\hbar \omega_p\sim r_s^{-1/2}$. 

\subsection{Static properties of uniform electron fluids}

\begin{figure}[t]
\hspace{-0.90cm}
\includegraphics[width=0.55\textwidth]{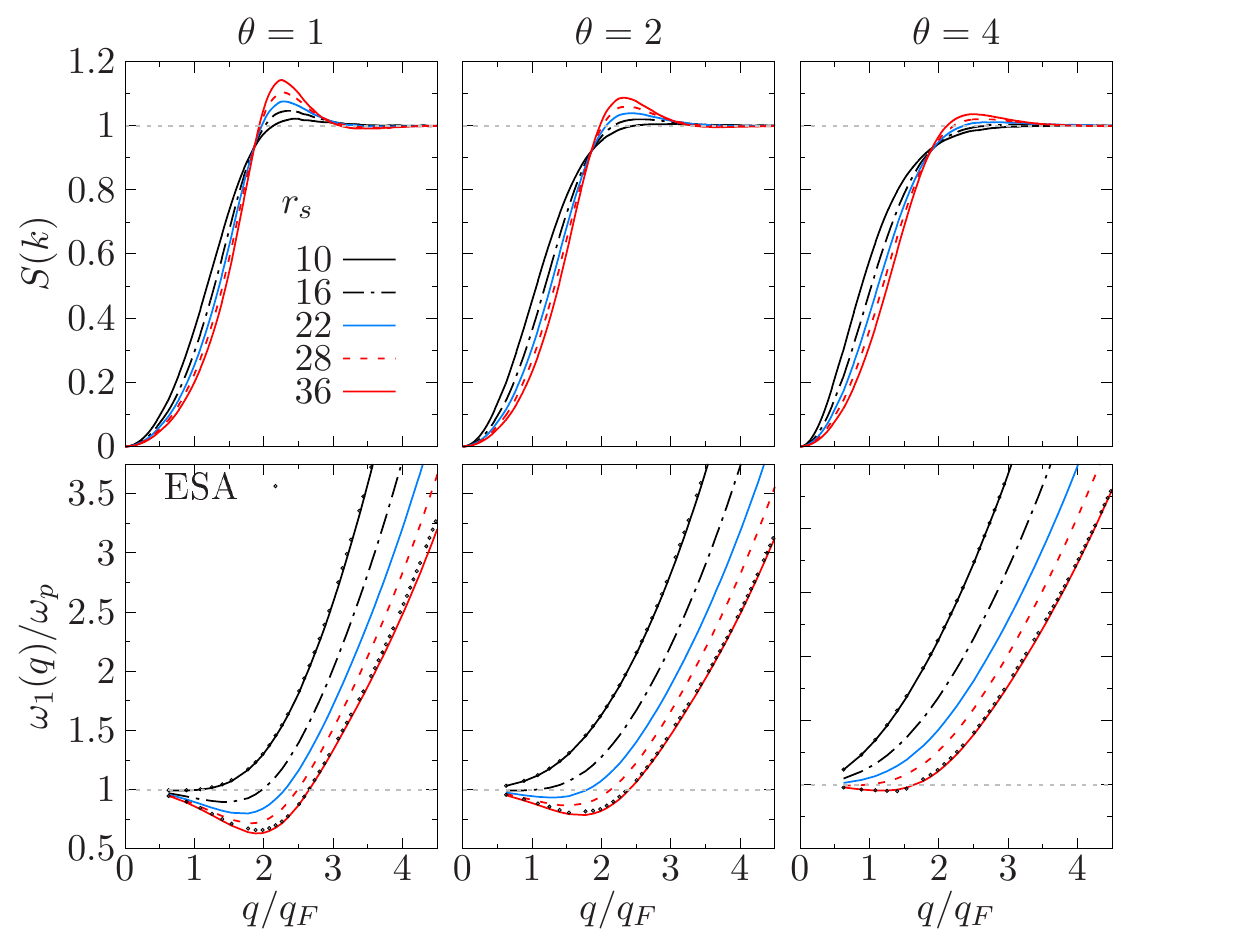}
\vspace{-0.6cm}
\caption{The static structure factor $S(q)=F(q,0)$ (obtained by the spline-interpolation, see Fig.~\ref{fig:SkSTLS}) and the first characteristic frequency, $\omega_1(q)=\sqrt{C_2(q)/C_0(q)}=\omega_p/\sqrt{C_0(q)}$
of UEG at $r_s=10,16,22,28,36$ and temperatures $\theta=1,2,4$. Solid dots correspond to $\omega_1(q)/\omega_p$ evaluated independently from the ESA model.
The presence of the second excitation branch in the spectrum (see Fig.~\ref{fig:DSF_Rs36}) is correlated with the observation of a local maximum in the SSF (i.e. $S(q)\geq 1$) for the wavenumbers $1.8 \leq q/q_F \leq 2.9$. 
}
\label{fig:staticRs36}
\end{figure}

The power moments, $C_{0}(q;r_{s},\theta)$ and $C_{4}(q;r_{s},\theta)$, along with the $f$-sum rule $C_{2}(r_{s})=\omega_{p}^{2}$, are the input of the 9MA model. This permits to express the DSF and the dynamical dielectric function in terms of the characteristic frequencies $%
\tilde{\omega}\left( q\right) =\{\omega _{1}\left( q\right) ,\omega
_{2}\left( q\right) ,\omega _{3}\left( q\right) ,\omega _{4}\left(
q\right) \}$ (see Eq.~\ref{oms}) with the additional parameters $\{\omega
_{3}\left( q\right) ,\omega _{4}\left( q\right) \}$ being determined at
given thermodynamic conditions from the first two characteristic frequencies
by the Shannon entropy maximization procedure or from the intermediate
scattering function as it is described below in Sec.\ref{SecReconstr}.

The results of our PIMC simulations for the low-density phase of the UEG are presented in Fig.~\ref{fig:staticRs36}, and clearly demonstrate the interplay of both correlations and temperature effects. The static structure factor (SSF), $S\left( q\right)$, and the first characteristic frequency 
\begin{equation}
\omega_{1}(q) =\left(C_{2}/C_{0}\right)^{1/2}=\omega_p\left(1-\epsilon^{-1}\left(q,0\right) \right)^{-1/2}  \label{w1a}
\end{equation}
directly related to the static inverse dielectric function (IDF), 
$\epsilon^{-1}\left( q,0\right)$, are shown as a function of the density parameter $r_s$ and the temperature.
It is important that Eq.~(\ref{w1a}) follows from the Kramers-Kronig relation for the IDF, which is a genuine response function. Thus the static IDF and the SSF are the real physical input quantities in our model.
In the lower panels, the characteristic frequency $\omega_1(q)$ is evaluated within the ESA model independently. These results are indicated by the solid dots (only for the lowest and highest $r_s$ values) and demonstrate a nice agreement with our present data.   

From Fig.~\ref{fig:staticRs36} we can unambiguously conclude that $\omega_{1}\left(q\right) <1$ in a certain wavenumber interval for $r_s\geq 16$ and $\theta \lesssim 2$, which is equivalent to negative values of the static dielectric function
\begin{equation}
\epsilon ^{-1}\left( k,0\right) =1/\epsilon \left( k,0\right) <0.
\label{negative}
\end{equation}%
for such conditions. The possibility and validity of this inequality is well-known as the {\it over-screening effect}, see~\cite{RevModPhys.53.81, PhysRevE.104.015202} and references therein. It is directly related to the analyticity of the direct dielectric function $\epsilon \left( q,z\right)$ in the upper half-plane of the complex
frequency plane, but this topic is beyond the scope of the present work. 

\subsection{Reconstruction of the higher-order moments $C_{6(8)}$}\label{SecReconstr}

The virtually unknown higher-order power moments $C_{6(8)}$ introduced above in Sec. II constitute a very important ingredient in the dynamical Nevanlinna parameter function. As it was demonstrated in Sec. IID, their reconstruction based on the maximization of the Shannon entropy functional leads to a nearly perfect agreement with the results based on the dynamical local field. The main advantage of the present approach is that we employ only a limited set of static characteristics $\{S(q),\chi(q,0)\}$. On the contrary, the DLFC reconstruction is mainly relied on a high-quality QMC data obtained for the density-density response function in the imaginary time. It is a peculiar decay of $F^{\text{PIMC}}(q,\tau_i)$, $(1\leq i\leq M)$, obtained with the fermionic PIMC,  that has allowed to reconstruct ab initio UEG DSF in the high and moderate density regime ($r_s\leq 10$).  The 9MA  demonstrates in this regime a similar accurate predictive power for the dynamical response, however, with much less computational effort.     

The main drawback of the Shannon-entropy approach, as it is already discussed in Sec. IIE, is the artificial smoothing of the sharp energy resonances, in particular, in the $q$ range spanned by the plasmon resonance. Additional information on the intermediate scattering function (ISF) available from the QMC data can be used to specify the results of the entropy approach for any wavenumber $q$. As a quantitative criterion, similar to the one used in the stochastic and the generic optimization techniques~\cite{mishchenko.prb2000,vitali.2010prb,filinov.2012pra}, we suggest to use to this end the relative deviation from the QMC data, 
\begin{eqnarray}
\delta F_r^{\text{trial}}(q)=\frac{\Delta \tau}{\beta}\sum\limits_{i=1}^M \frac{|F^{\text{trail}}(q,\tau_i)-F^{\text{QMC}}(q,\tau_i)|}{F^{\text{QMC}}(q,\tau_i)} 
\end{eqnarray}
integrated along the imaginary time $0\leq \hbar \tau_i\leq \hbar \beta$, with $M$ being the number of high-temperatures propagators and $\Delta \tau=\tau_{i+1}-\tau_i=\beta/M$. 

In addition, we introduce a natural measure of the statistical noise present in the QMC data 
\begin{eqnarray}
\delta F_r^\text{QMC}(q)=\frac{\Delta \tau}{\beta}\sum\limits_{i=1}^M \frac{\delta F^{\text{QMC}}(q,\tau_i)}{F^{\text{QMC}}(q,\tau_i)},
\label{err_QMC}
\end{eqnarray}
where $\delta F^{\text{QMC(PIMC)}}(q,\tau_i)$ is the statistical uncertainty in the evaluation of ISF. 

\begin{figure}[t]
\hspace{-0.90cm}
\includegraphics[width=0.51\textwidth]{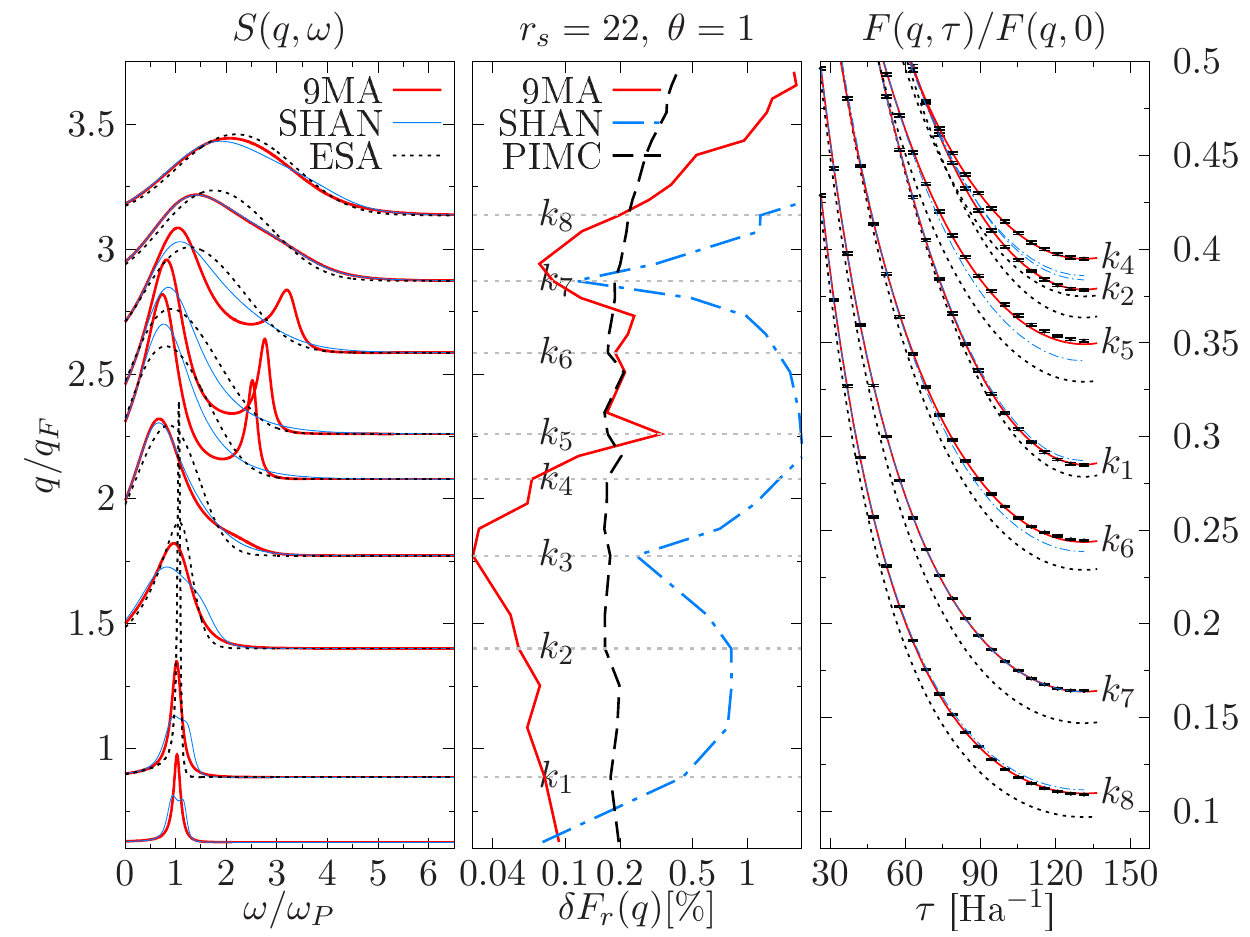}
\vspace{-0.2cm}
\caption{(from left to right) The dynamic structure factor $S(q_i,\omega)$ for $r_s=22$ ($\theta=1$) and selected wavenumbers $k_i=q_i/q_F$ from the three models: the ESA and the method of moments with the frequencies $\omega_{3(4)}$ (moments $C_6, C_8$) reconstructed with the Shannon entropy ("SHAN") and as the fit to the intermediate scattering function $F(q,\tau)$ ("9MA") [the optimized solution in $\omega_{3(4)}$], along the relative deviation measure $\delta F_r(q)$ [in percentage points]  of two of these models ("SHAN", "9MA") from $F^{\text{PIMC}}(q,\tau)$. The dashed black line "PIMC" stands for the statistical uncertainty in the PIMC data, Eq.~(\ref{err_QMC}). The normalized ISF from the three models ("ESA","SHAN","9MA") vs. ab initio PIMC data are represented by black symbols with error bars. The ISF is shown only up to $\tau=\beta/2$ due to the symmetry, $F(q,\tau)=F(q,\beta-\tau)$, provided by the DSF detailed balance condition, $S(q,-\omega)=e^{-\beta \hbar \omega} S(q,\omega)$.}
\label{fig:DSFrs22b1_optim}
\end{figure}

For the wavenumbers $q$ such that the Shannon-entropy-based solution leads to the reconstructed ISF, 
i.e. $S^{\text{trial}}(q,\omega)\Rightarrow F^{\text{trial}}(q,\omega)$, which satisfies the criterion 
\begin{eqnarray}
\delta F_r^{\text{trial}}(q)\lesssim \delta F_r^{\text{QMC}}(q),
\label{dFtau_cond}
\end {eqnarray}
this solution can be accepted as a plausible physical solution, which in addition satisfies the set of involved power moments exactly. In the $q$ segment where such condition is violated, a refinement of a trial entropy-based solution is necessary. This approach has been successfully used in the reconstruction of the plasmon feature as presented in Fig.~\ref{fig:Ftau}, where the higher-order moments $C_{6(8)}$ (or $\omega_{3(4)}$) were used as the fitting parameters to satisfy the acceptance criterion~(\ref{dFtau_cond}).

In the analysis of the low-density regime ($16 \leq r_s\leq 36$), discussed below in detail in Secs. IIIC and IIID, we have followed a similar strategy:
\begin{enumerate}
    \item The Shannon EM solution, $S^{\text{SH}}(q,\omega)$, and $\{\omega^{\text{SH}}_{3(4)}\}$ is obtained in the full range of wavenumbers; 
    \item  The trial entropy-based solution for the ISF, i.e. $F^{\text{SHAN}}(q,\tau)$, is constructed and verified against the acceptance condition~(\ref{dFtau_cond});
    \item The Shannon frequencies are considered as the initial parameters, $\omega^{(0)}_{3(4)}=\omega^{\text{SH}}_{3(4)}(q)$ for the solution of the optimization problem
\begin{eqnarray}
 &&\min_{\{\omega_{3(4)}(q)\}}{\delta F_r^{\text{trial}}(q;\omega_{3},\omega_4)},\\ &&\frac{\partial F_r^{\text{trial}}(q;\omega_{3},\omega_4)}{\partial \omega_{3(4)}}=0
\end{eqnarray}
via the Newton-Raphson method. In the sequence of iterations, $\{\omega^{(n-1)}_{3(4)}\rightarrow \omega^{(n)}_{3(4)}\}$, at every step $n$ the corresponding quantities are reevaluated:
\begin{enumerate}
 \item $\{\omega_{1(2)};\omega^{(n)}_{3(4)}\}\rightarrow S^{(n)}(q,\omega_{1(2)};
 \omega^{(n)}_{3(4)})\rightarrow F^{(n)}(q,\tau;\omega_{1(2)},\omega^{(n)}_{3(4)} )\rightarrow  \delta F_r^{(n)}(q;\omega_{1(2)},\omega^{(n)}_{3(4)})$.
\end{enumerate}

\item For the wavenumber values with $\delta F_r^{\text{SH}}(q)> \delta F_r^{QMC}(q)$ and $\delta F_r^{\text{trial}}(q;\omega_{3},\omega_4) < \delta F_r^{\text{SH}}(q;\omega^{\text{SH}}_{3},\omega^{\text{SH}}_4)$ the initial Shannon frequencies are substituted by the optimised solutions.
\end{enumerate}

An example of the optimization procedure for $r_s=22$ and $\theta=1$ is presented in Fig.~\ref{fig:DSFrs22b1_optim}. The left-hand panel shows three model DSFs for the selected values of $q$. The 9MA solution with the Shannon and the optimized frequencies (denoted as "SHAN" and "9MA") are shown along with the ESA solution. For each case the corresponding ISF was evaluated (see the right-hand panel) and the qualifying deviation measure $\delta F_r(q)$ (the central panel) was estimated to confirm the acceptance condition~(\ref{dFtau_cond}). The measure of the statistical noise $\delta F_r^{\text{PIMC}}(q)$ in the PIMC data is demonstrated by the dashed black line (the central panel). As one can see, among three models only the 9MA solution with the dynamical Nevanlinna parameter function and the optimized frequencies $\omega_{3(4)}$ satisfies the acceptance condition for all $q \leq 3.2 q_F$, and predicts new energy resonances around $q\sim 2.2 q_F$ (for a full DSF see Sec. IIIC). The corresponding $q$ segment with this new feature is close to the position of the broad maximum in the SSF, see Fig.~\ref{fig:staticRs36}. Both ESA and SHAN models fail to predict a high-energy eigenmode for the selected wavenumbers ($q_4=2.08 q_F$, $q_5=2.17 q_F$ and $q_6=2.51 q_F$) and reproduce a broad distribution with a high-frequency shoulder. Next, we observe that the low-frequency DSF maximum in the SHAN and the 9MA solutions nearly coincide, while the ESA peak position is always shifted to higher frequencies. The same trend was already observed in the moderate density regime ($r_s=6; 10$, see Fig.~\ref{fig:skdynb1}), where both solutions with the dynamical correlations (SHAN and DLFC) demonstrate a very good agreement and a {\it redshift} with respect to the predictions of the ESA model. 

\begin{figure}[t]
\hspace{-0.90cm}
\includegraphics[width=0.51\textwidth]{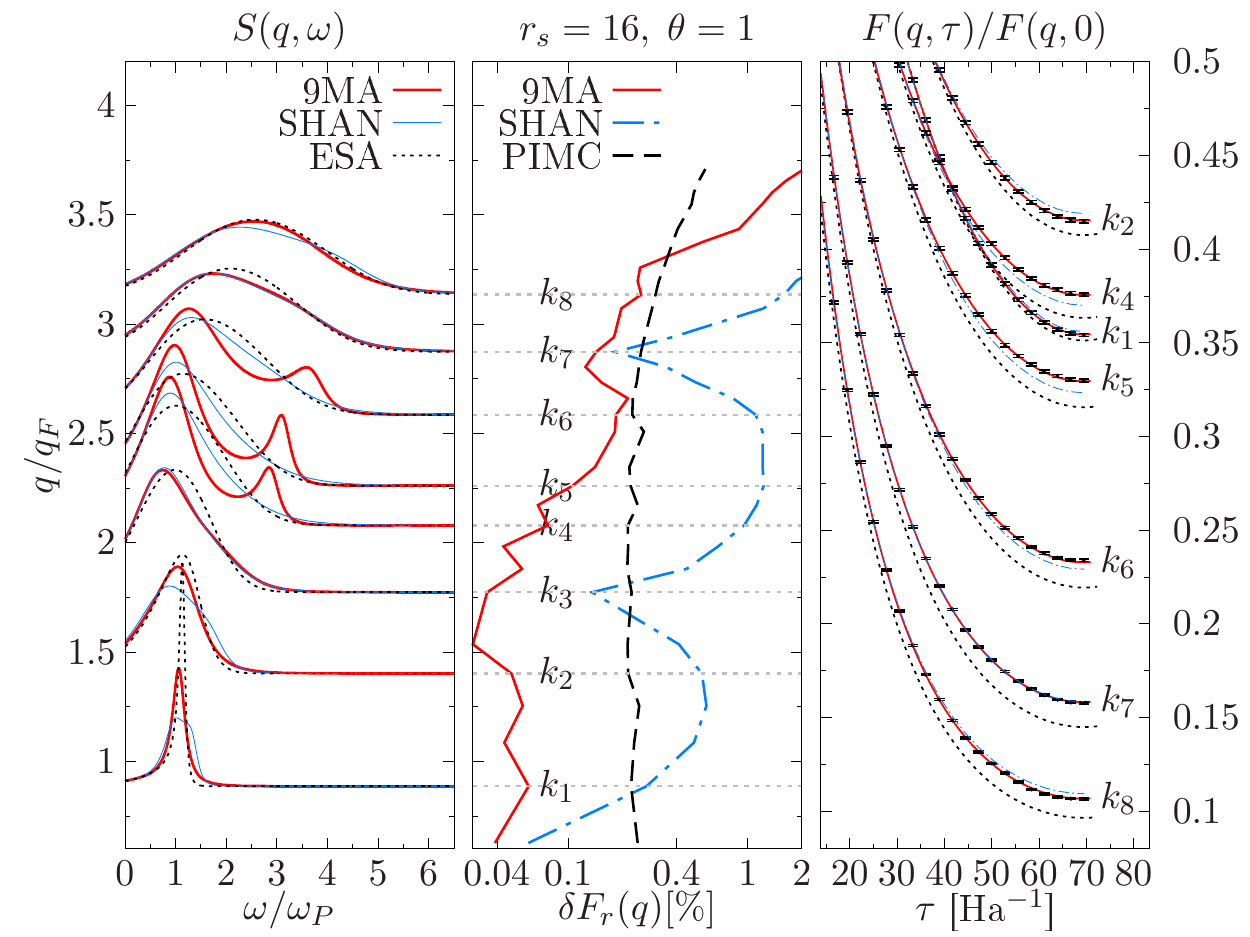}
\vspace{-0.2cm}
\caption{As in Fig.~\ref{fig:DSFrs22b1_optim} but for $r_s=16$ and $\theta=1$.}
\label{fig:DSFrs16b1_optim}
\end{figure}

\begin{figure}[t]
\hspace{-0.90cm}
\includegraphics[width=0.51\textwidth]{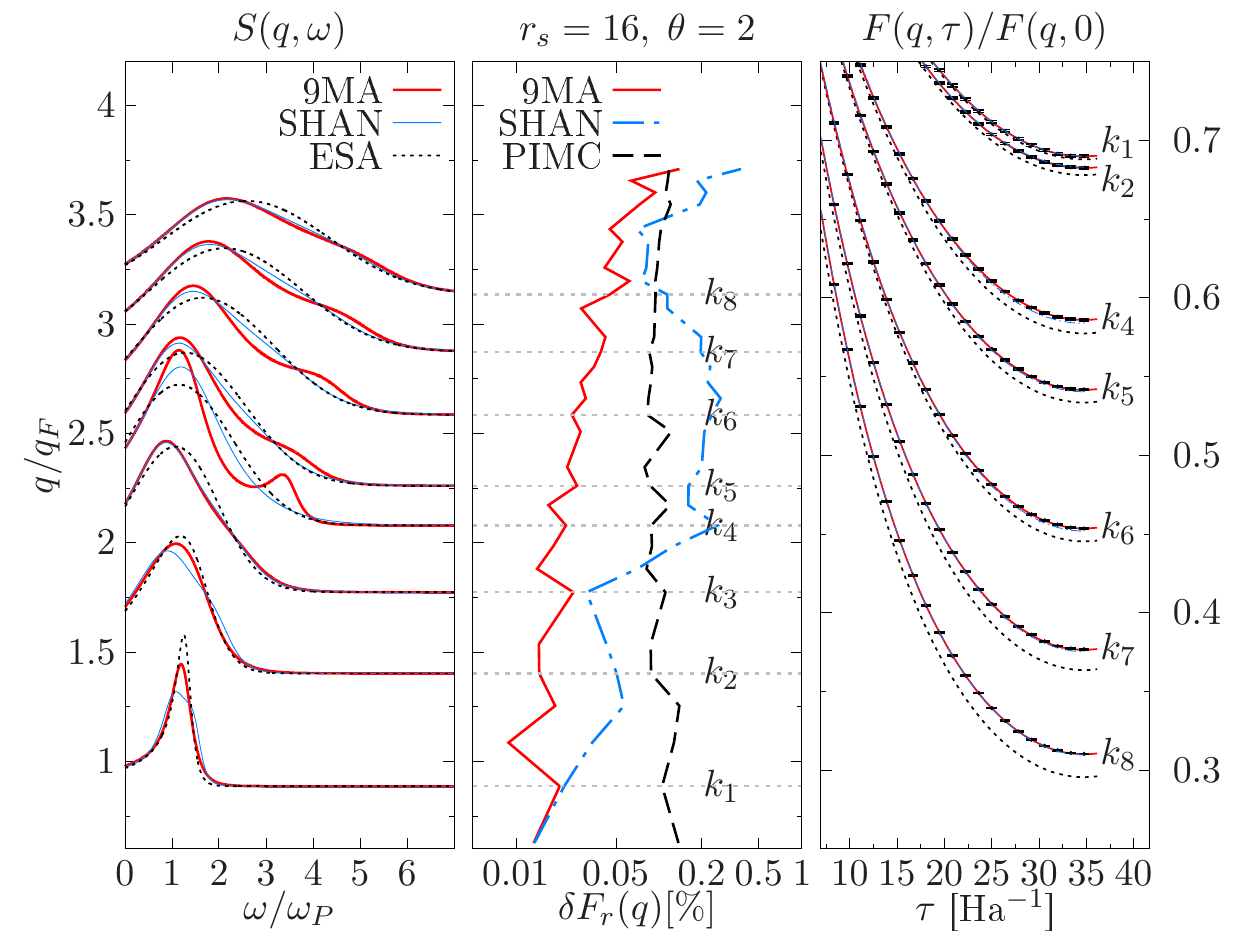}
\vspace{-0.2cm}
\caption{As in Fig.~\ref{fig:DSFrs22b1_optim} but for $r_s=16$ and $\theta=2$.}
\label{fig:DSFrs16b2_optim}
\end{figure}

This fact is reflected in the asymptotic behaviour of the intermediate scattering function $F(q,\tau)$ as $\tau \rightarrow \beta/2$, see the right-hand panel in Fig.~\ref{fig:DSFrs22b1_optim}. Here, we observe that the 9MA and SHAN solutions are in a very good agreement with ab initio PIMC data (symbols with the error bars), while the ESA ISF, $F^{\text{ESA}}(q,\tau)$ (dotted black curves), demonstrate systematic and significant deviations with some acceptable agreement with the PIMC data being achieved only for the smallest wavenumbers $\{q_1,q_2\}$ when only a single plasmon resonance ($\omega(q)\sim \omega_p$) dominates in the full spectral density. Notice, however, that even in this case the plasmon width (decrement) is underestimated by the ESA model and leads to small but noticeable deviations in the asymptotic value $F^{\text{ESA}}(q,\beta/2)$. Similar observations apply to the SHAN solution at $q_2$. Here, in contrast, the plasmon feature is smoothed by the maximization of the entropy functional and leads to the overestimation of the plasmon decrement against the optimized solution: compare the DSF plots $S^{\text{SH}}(q_2,\omega)$,  $S^{\text{9MA}}(q_2,\omega)$ on the left-hand panel.  

In summary, we can qualify different trial DSF solutions based on the deviation measure introduced above and presented in the central panel of Fig.~\ref{fig:DSFrs22b1_optim}. The deviations $\delta F^{\text{ESA}}_r(q)$ exceed $2\%$ and are not shown. The SHAN model allows to reduce the deviation measure, $\delta F^{\text{SH}}_r(q)$, by an order of magnitude but it still significantly exceeds the upper bound specified by the statistical noise, $\delta F_r^{QMC}(q)$. Hence, only the optimized solution 9MA is acceptable at these conditions.

Similar analysis has been performed for  $\{r_s=16; 22; 28; 36\}$ and $\{\theta=1; 1.5; 2; 4; 8\}$. More  examples are presented in Figs.~\ref{fig:DSFrs16b1_optim},\ref{fig:DSFrs16b2_optim} and lead us to several important conclusions. First, the double-peak DSF structure is reproduced at all analysed densities ($16 \leq r_s\leq 36$) and low temperatures ($\theta\lesssim 2$) but only in a finite range of wavenumbers, $1.77 \lesssim q/q_F\lesssim 2.9$. Both ESA and SHAN models are missing this important spectral feature and violate in this part of the spectrum the acceptance condition~(\ref{dFtau_cond}). The deviation measure of the entropy-based solution (SHAN) is significantly reduced with increasing temperature so that at $\theta\gtrsim 4$ it becomes comparable to the optimized solution, i.e $\delta F^{\text{SHAN}}_r(q) \sim \delta F^{\text{9MA}}_r(q)$. Even, at $\theta=2$, as it is demonstrated in the central and the right-hand panels of Fig.~\ref{fig:DSFrs16b2_optim}, the SHAN solution already reproduces the ISF, $F^{\text{PIMC}}(q,\tau)$, within the error bars, except for the interval $k_3< k< k_8$, where some reminder of the second shifted mode is still visible. Notice that the integrated deviation measure, $\delta F^{\text{SH}}_r(q)$, at this temperature does not exceed $0.2\%$ while at $\theta=1$ it might reach $1\%$,  (Fig.~\ref{fig:DSFrs16b1_optim}). 
The suppression of the high-frequency resonances with $\theta$, c.f. Figs.~\ref{fig:DSFrs16b1_optim},\ref{fig:DSFrs16b2_optim} observed here is analysed in detail in Sec. IIID.  

Finally, the above analysis supports our previous conclusion with respect to the applicability of the Shannon-entropy approach at high and moderate densities ($r_s\leq 10$). Once the interaction and decay processes of the quasiparticle excitations result in a smooth and slow varying spectral density, the entropy principle applies and already leads to an optimized DSF form related to a physically relevant solution. Moreover, the  entropy maximization permits to reconstruct a physically reliable model of the dynamical Nevanlinna function using the compact representation based on only two optimization parameters $\{\omega_{3(4)}(q)\}$. This fact is proved by the present detailed analysis and, in our opinion, has a clear advantage over the complex and not physically transparent representation of the DLFC function of ~\cite{dornheim.2018prl,groth.2019prb} which followed the idea of Dabrowski~\cite{PhysRevB.34.4989} motivated by exact DLFC limiting forms by introducing an "extended" Padé-type expression for the imaginary part of the DLFC with six "random" parameters.

\subsection{Dynamical structure factor: observation of the second excitation branch and temperature effects}\label{DSFrs36}

\begin{figure}[t]
\hspace{-0.90cm}
\includegraphics[width=0.51\textwidth]{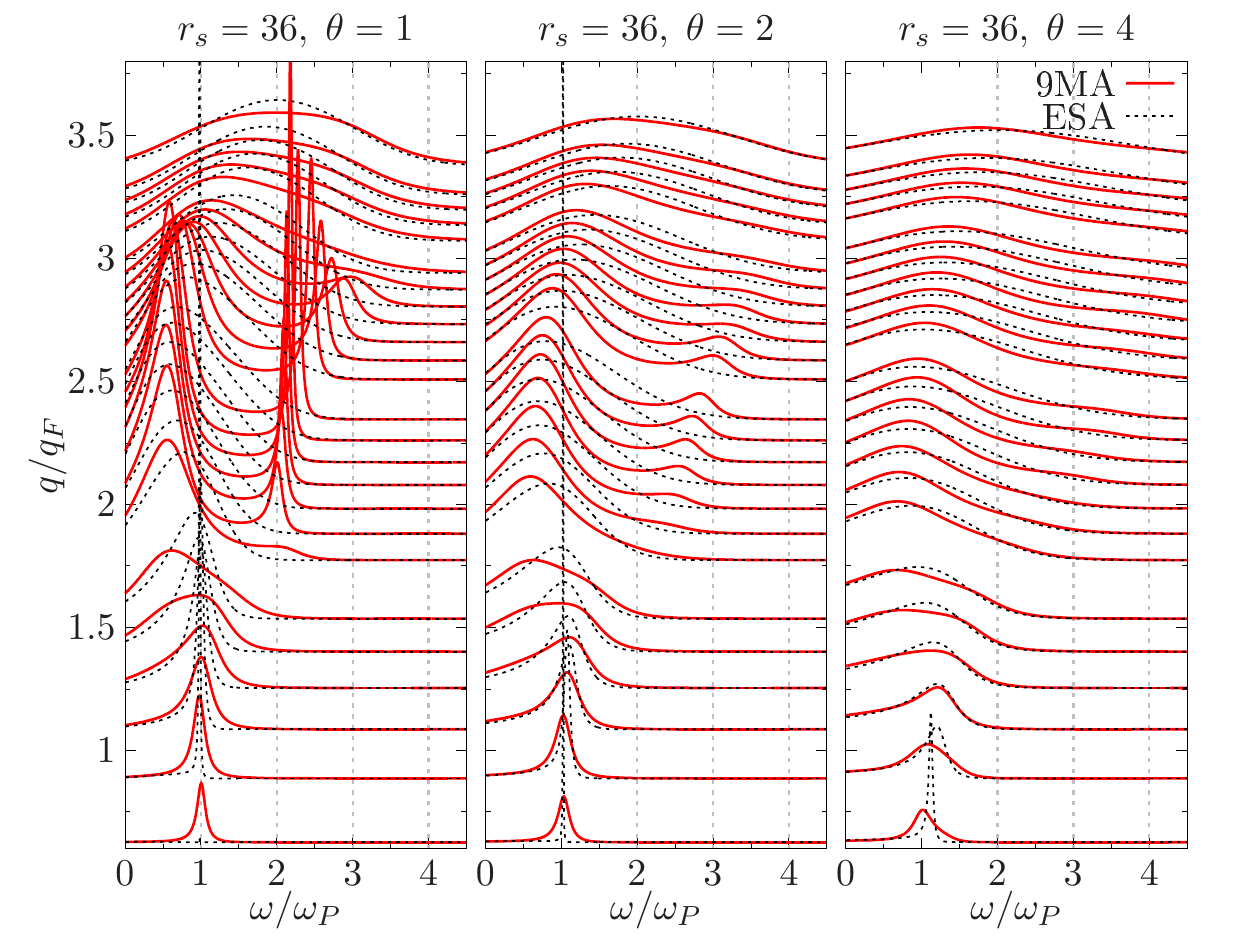}
\vspace{-0.2cm}
\caption{The dynamic structure factor $S(q,\omega)$ for $r_s=36$ and $\theta=1; 2; 4$. A clear signature of the second harmonic is observed for $\theta \lesssim 2$. The temperature increase results in the smoothing of this feature.}
\label{fig:DSF_Rs36}
\end{figure}

\begin{figure}[t]
\hspace{-0.90cm}
\includegraphics[width=0.51\textwidth]{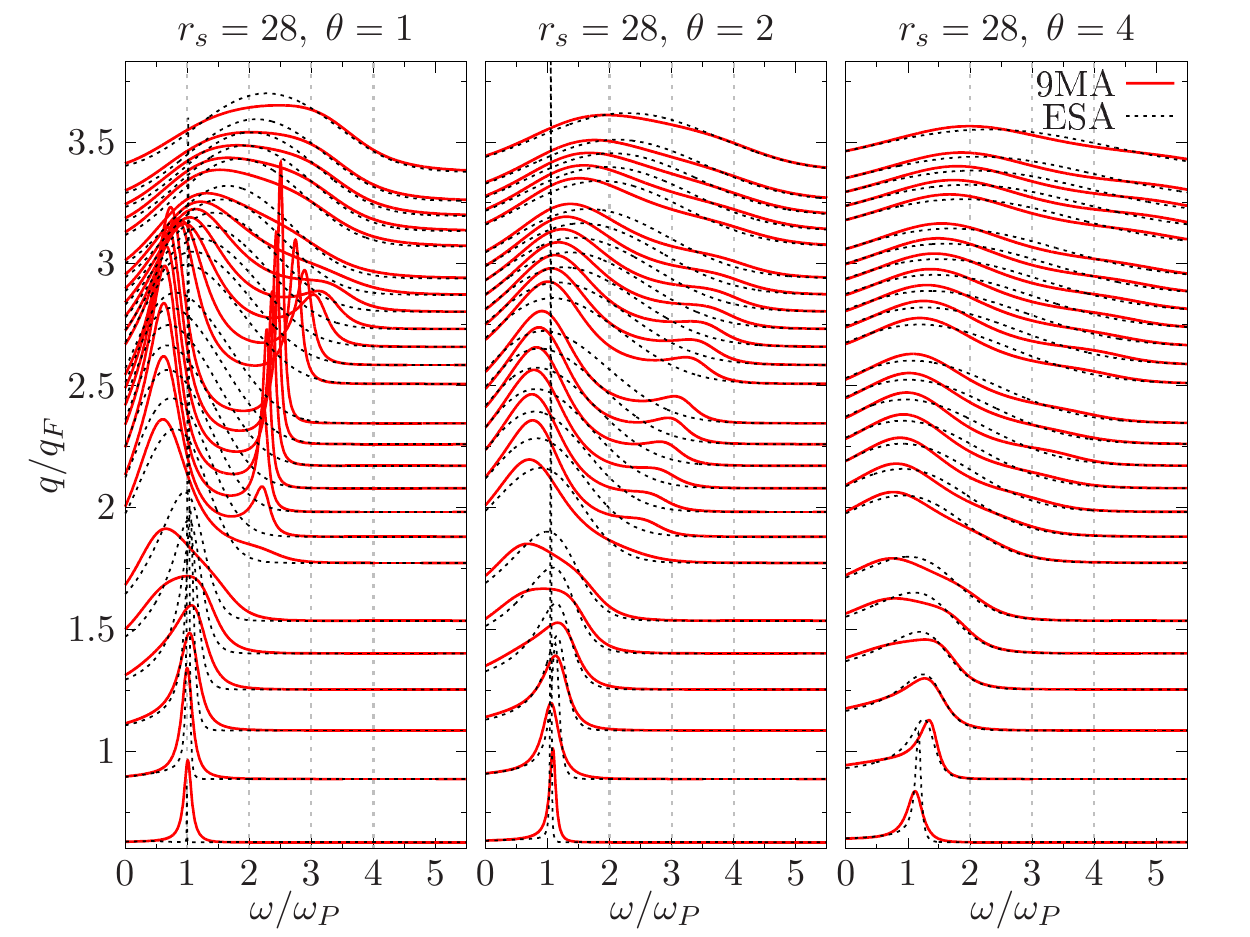}
\vspace{-0.2cm}
\caption{The dynamic structure factor $S(q,\omega)$ for $r_s=28$ and $\theta=1; 2; 4$. A clear signature of the second harmonic is observed for $\theta \lesssim 2$. The temperature increase effectively eliminates this feature.}
\label{fig:DSF_Rs28}
\end{figure}

\begin{figure}[t]
\hspace{-0.90cm}
\includegraphics[width=0.51\textwidth]{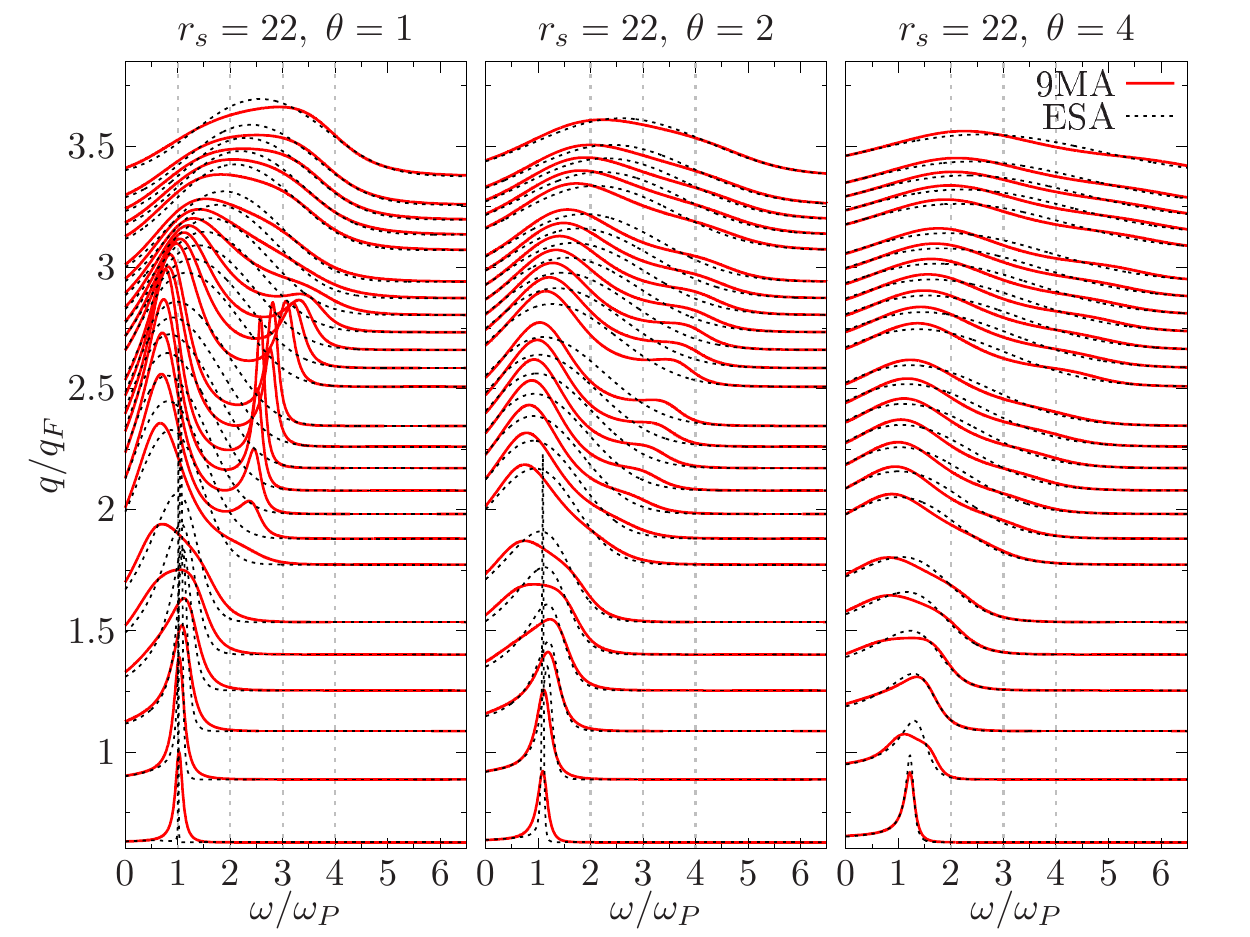}
\vspace{-0.2cm}
\caption{The dynamic structure factor $S(q,\omega)$ for $r_s=22$ and $\theta=1; 2; 4$. A clear signature of the second harmonic is observed for $\theta \lesssim 2$. The influence of the temperature increase is confirmed.}
\label{fig:DSF_Rs22}
\end{figure}

\begin{figure}[t]
\hspace{-0.90cm}
\includegraphics[width=0.51\textwidth]{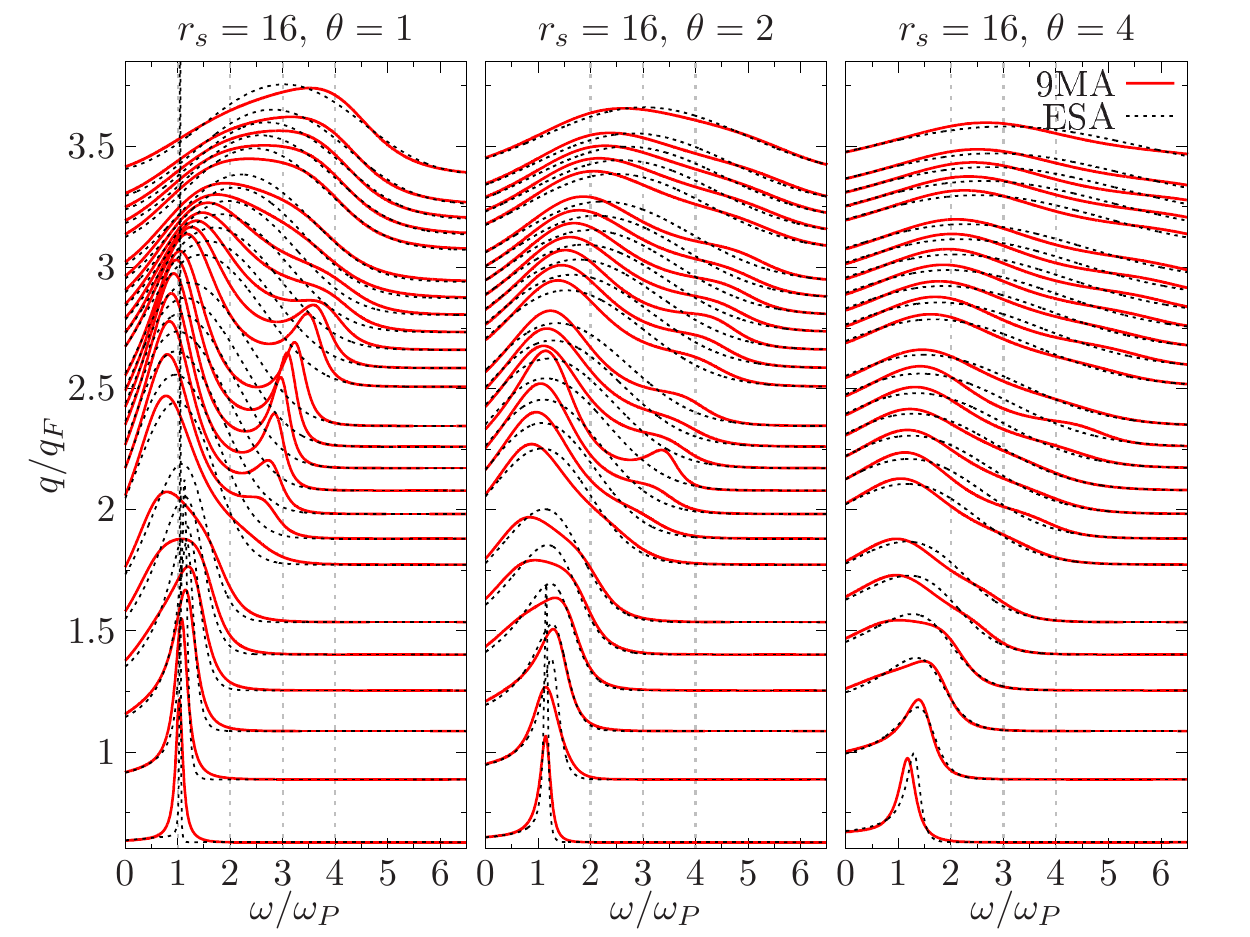}
\vspace{-0.2cm}
\caption{The dynamic structure factor $S(q,\omega)$ for $r_s=16$ and $\theta=1; 2; 4$. A clear signature of the second harmonic is observed for $\theta \lesssim 2$. The temperature increase leads to the smearing out of the second harmonic feature. }
\label{fig:DSF_Rs16}
\end{figure}

Here we provide some graphical representations of the UEG excitation spectrum in the low-density regime ($16 \leq r_s \leq 36$) based on the accurate reconstruction recipe presented in the previous section. Three temperature cases are shown in  Figs.~\ref{fig:DSF_Rs36}, \ref{fig:DSF_Rs28}, \ref{fig:DSF_Rs22}, \ref{fig:DSF_Rs16} with a pronounced emergence of the high-frequency mode starting at $q\gtrsim 1.77 q_F$ and $\omega \gtrsim 2 \omega_p$, which, at first sight, can be attributed to the {\it double plasmon excitation}. Comparing different density cases, the sharpest energy resonances are observed at the lowest density $r_s=36$ and the lowest physical temperature, $\theta=T/T_F=1$, due to the scaling $T_F\sim r_s^{-2}$. By decreasing the electron gas density from $r_s=36$ ($\theta=1$) to $r_s=16$ ($\theta=1$) we demonstrate a systematic shift of the high energy branch to higher frequencies along with the damping enhancement. For all density cases at $\theta=1$ the upper mode can be observed only up to $q\sim 2.9 q_F$, and for larger wavenumber values it transforms the DSF into a broad distribution with a single maximum.   
Simultaneously, in the same wavenumber interval ($1.77 \lesssim q/q_F \lesssim 2.9$) a well defined low-frequency mode is present possessing a {\it roton}-like feature in the dispersion curve. Similar effect has already been observed at higher densities, cf. $r_s=10$ in Fig.~\ref{fig:modes}.  

Next, the central and right-hand panels in Figs.~\ref{fig:DSF_Rs36},\ref{fig:DSF_Rs28},\ref{fig:DSF_Rs22},\ref{fig:DSF_Rs16} demonstrate the redistribution of the spectral weight and the damping of both modes when the temperature increases. 
At $\theta=2$ there is some reminiscence of the second branch, while at $\theta=4$ we can only observe a high-frequency shoulder observed previously for  $r_s=6$ and $r_s=10$ (cf. Fig.~\ref{fig:skdynb1}). Thus, when $\theta=4$ both modes become overdamped and cannot be well separated in the DSF. This result is found to be in full agreement with our previous discussion in Sec.~\ref{interAnal}.

\begin{figure}[t]
\hspace{-0.90cm}
\includegraphics[width=0.54\textwidth]{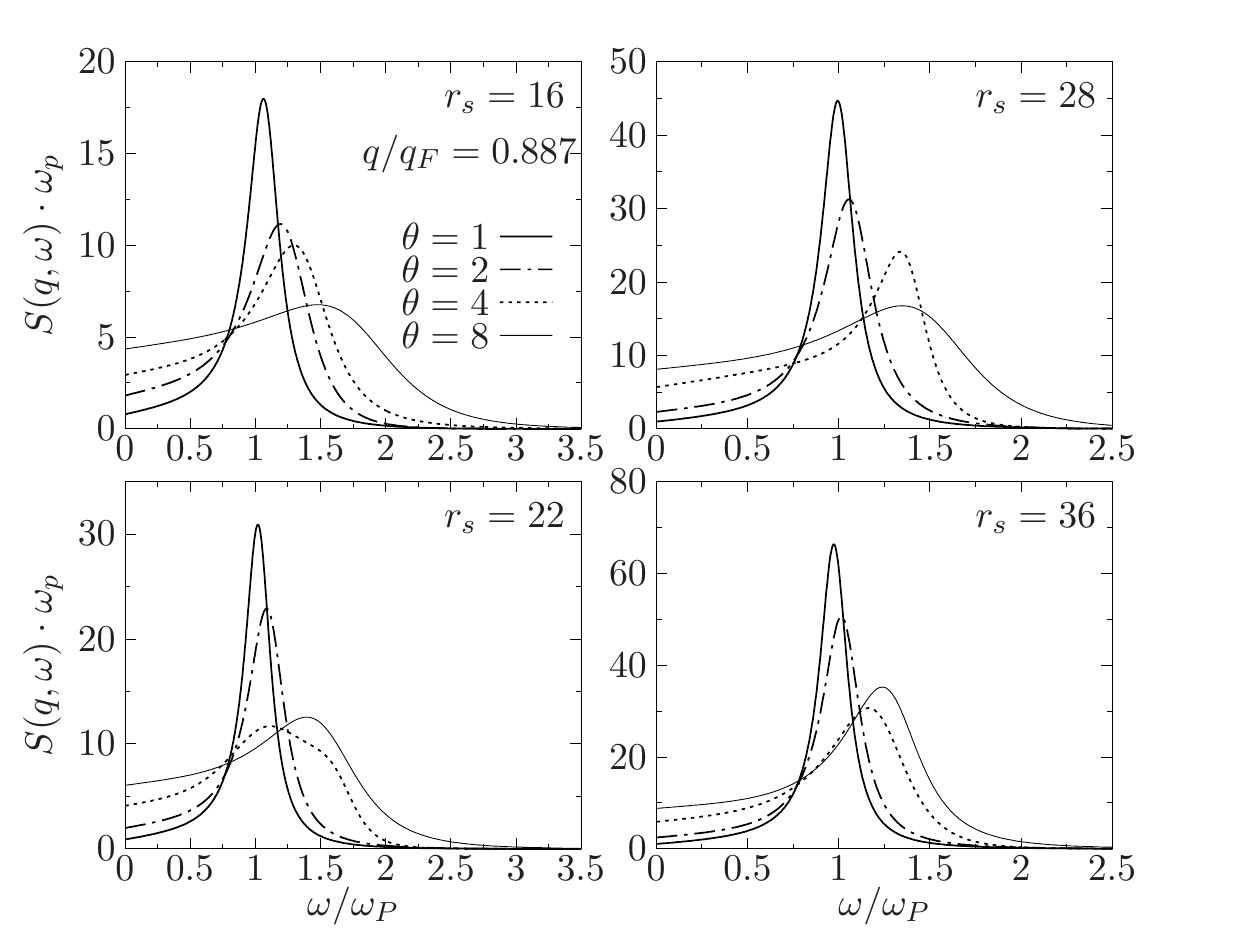}
\vspace{-0.6cm}
\caption{The $\theta$-dependence ($\theta=1; 2; 4; 8$) of $S(q,\omega)$ for $r_s=16; 22; 28; 36$. The wavenumber $q=0.887 q_F$ corresponds to the plasmon region.}
\label{fig:skdynTq1}
\end{figure}

\begin{figure}[t]
\hspace{-0.90cm}
\includegraphics[width=0.54\textwidth]{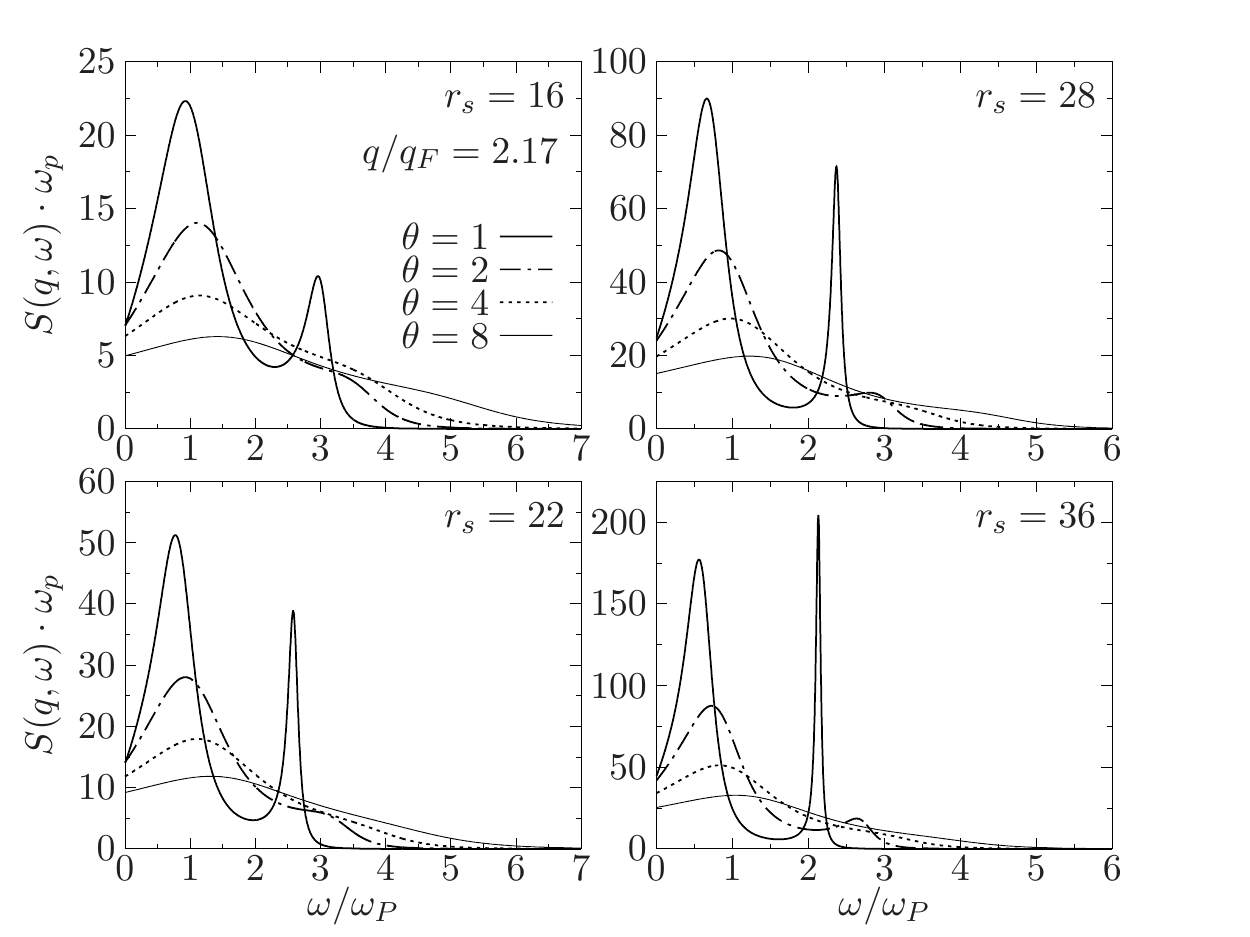}
\vspace{-0.6cm}
\caption{The $\theta$-dependence ($\theta=1; 2; 4; 8$) of $S(q,\omega)$ for $r_s=16; 22; 28; 36$. The wavenumber $q=2.17 q_F$  corresponds to the roton region.}
\label{fig:skdynTq10}
\end{figure}

\begin{figure}[t]
\hspace{-0.90cm}
\includegraphics[width=0.54\textwidth]{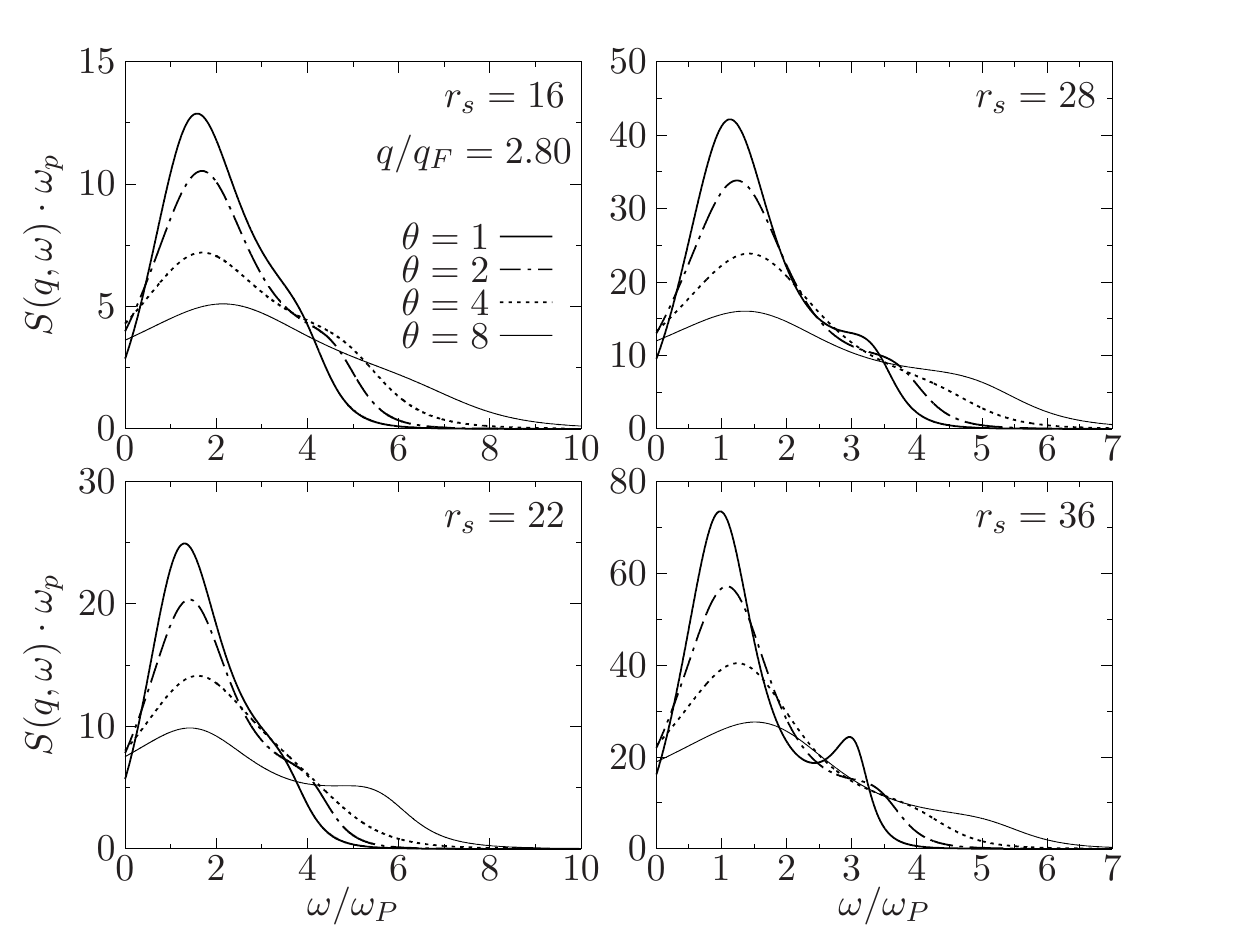}
\vspace{-0.6cm}
\caption{The $\theta$-dependence ($\theta=1; 2; 4; 8$) of $S(q,\omega)$ for $r_s=16; 22; 28; 36$. 
The specified wavenumber $q=2.80 q_F$ is beyond the roton region.}
\label{fig:skdynTq17}
\end{figure}

The explicit temperature dependence of the DSF at three different wavenumber values corresponding to the {\it plasmon}, {\it roton} and {\it beyond the roton} segments of the spectrum  is presented in Figs.~\ref{fig:skdynTq1},~\ref{fig:skdynTq10},~\ref{fig:skdynTq17}.
The two-mode structure is clearly seen in the spectrum within the roton segment which evolves into the pattern with the high-frequency shoulder when the higher mode becomes strongly overdamped .

\subsection{Dispersion relations: confirmation of the high-energy quasiparticle branch}\label{DSFDis}

\begin{figure}[t]
\hspace{-0.9cm}
\includegraphics[width=0.515\textwidth]{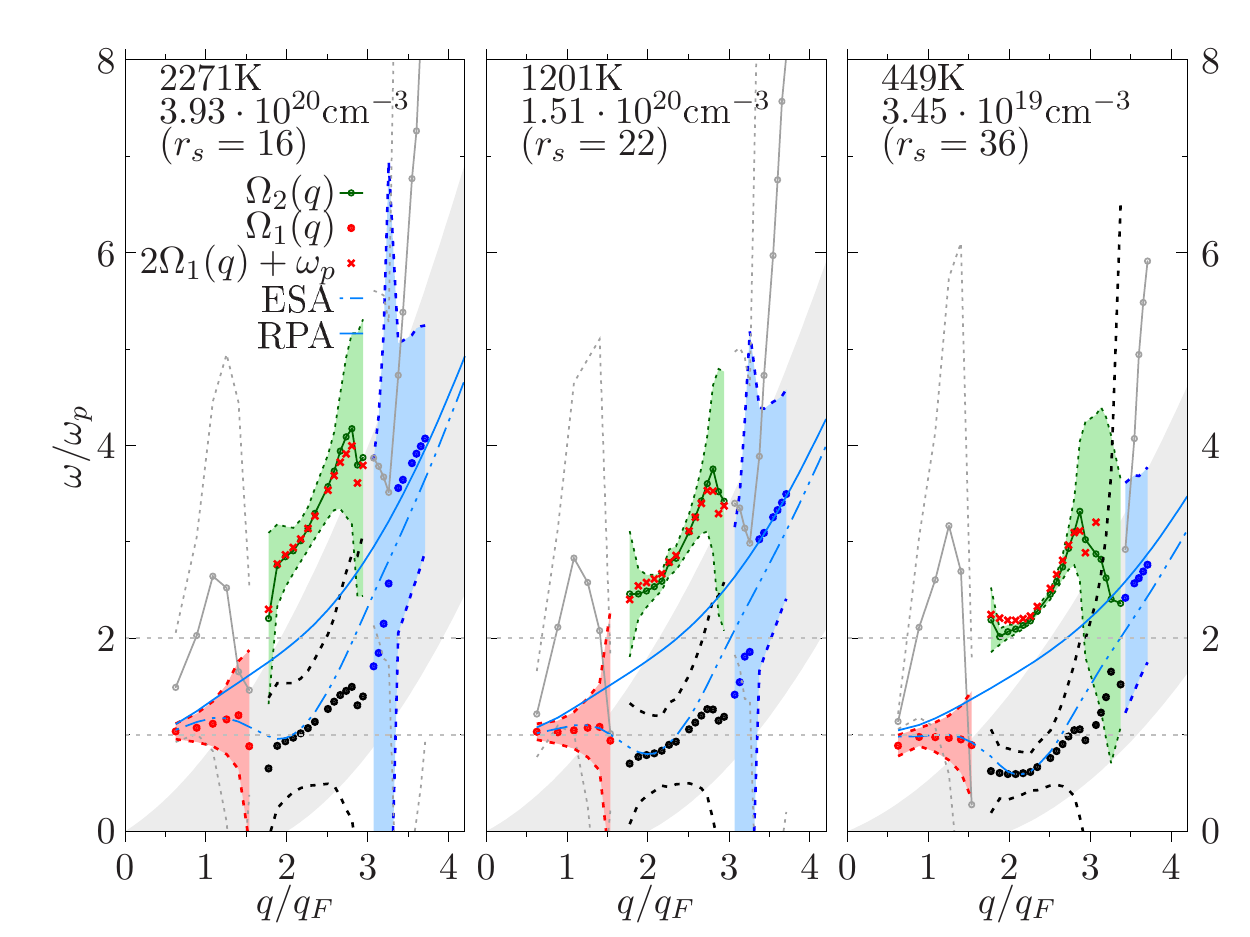}
\vspace{-0.3cm}
\caption{The wavenumber dependence of the solutions of the explicit dispersion equation, $\epsilon \left( q,z\right)=0$. Simulation parameters: $\theta=1$ and $r_s=16,22$ and $36$. Absolute physical temperatures/densities for UEG are included in the figure panels.
Two sets of symbols denote the modes in the three (plasmon/roton/free particle) wavenumber segments: $\Omega_{1}(q)$ (red/black/blue dots) and $\Omega_{2}(q)$ (grey/green/grey dots). The dashed lines standing for $\Omega_{1(2)}(q)-\Delta \Omega_{1(2)}(q)$ and $\Omega_{1(2)}(q)+\Delta \Omega_{1(2)}(q)$, where $\Delta \Omega_{1(2)}(q)$ are the decrements, represent the widths of the spectral lines. The position of the DSF maxima in the RPA/ESA solutions (solid blue/dashed black lines) are included for comparison. The red crosses, $2 \Omega_1(q)+\omega_p$, represent the combination of three quasiparticle excitations within the {\it roton segment} ($1.5 \leq q/q_F \leq 3$): two quasiparticle with a roton dispersion, $2 \Omega_1(q)$, plus a plasmon with the frequency $\omega_p$.}
\label{fig:modesRs36b1}
\end{figure}

\begin{figure}[t]
\hspace{-0.9cm}
\includegraphics[width=0.515\textwidth]{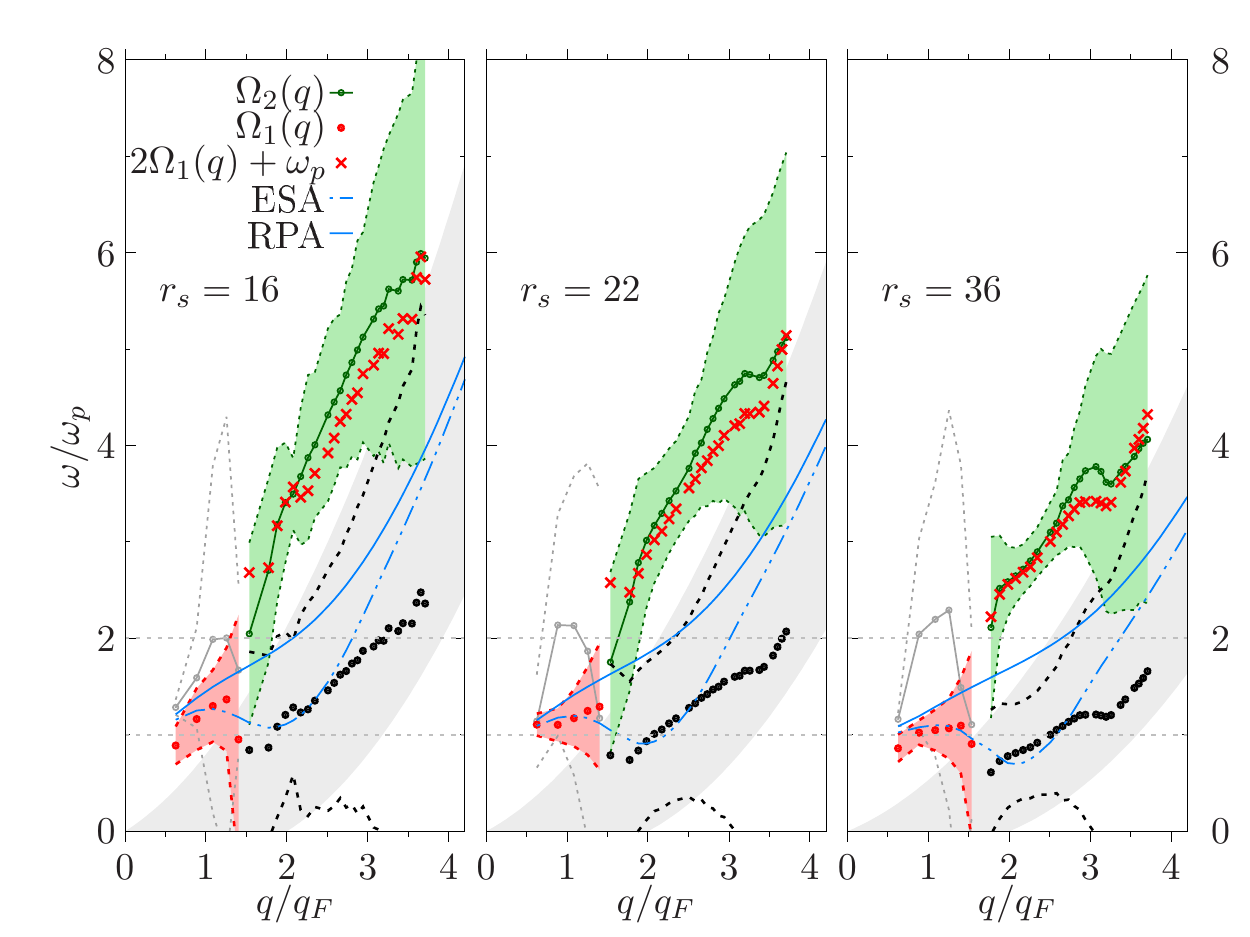}
\vspace{-0.3cm}
\caption{As in Fig.~\ref{fig:modesRs36b1} but for $\theta=2$.}
\label{fig:modesRs36b2}
\end{figure}

\begin{figure}[t]
\hspace{-0.9cm}
\includegraphics[width=0.515\textwidth]{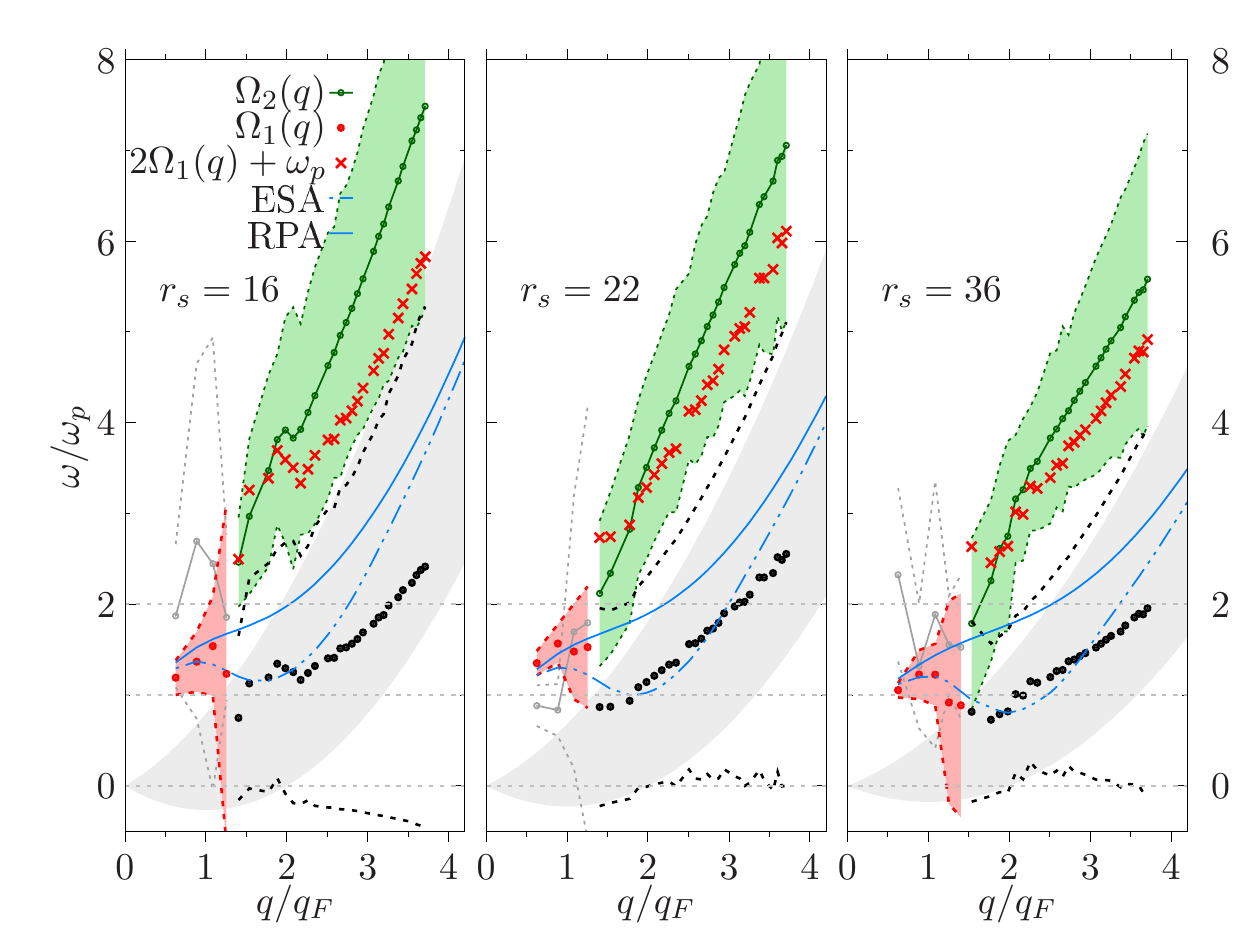}
\vspace{-0.3cm}
\caption{As in Fig.~\ref{fig:modesRs36b1} but for $\theta=4$. The frequencies $\omega_{3(4)}(q)$ are found within the 9MA model as a best fit to $F(q,\tau)$. Notice that due to the weak dependence of $F(q,\tau)$ on slight variations in $\omega_{3(4)}(q)$ at temperatures $\theta\gtrsim 4$ the solutions of the dispersion equation, $\Omega_{1(2)}(q)$ have uncertainties similar to those of the input values $\omega_{3(4)}(q)$ and, hence, we obtain non-smooth dispersion curves (mostly in the plasmon region). This problem is not present at lower temperatures ($\theta=1;2$) and can be partially subdued by the employment of the Shannon frequencies $\omega^{\text{SHAN}}_{3(4)}(q)$ resolved using the entropy maximization principle being applicable beyond the plasmon region ($q> q_c$) and higher temperatures as discussed in Sec.~\ref{Sec2C}. The improved dispersion is presented in Fig.~\ref{fig:modesRs36b4SHAN}.}
\label{fig:modesRs36b4}
\end{figure}

In this subsection we present our results with respect to the solutions of the explicit dispersion equation, 
$\epsilon \left( q,z\right)=0$, $z$ being the complex frequency. The dispersion relations of both characteristic modes, $\Omega_{1(2)}(q)$, and their respective decrements, $\Delta \Omega_{1(2)}(q)$, are found within the 9MA approximation and are displayed in Figs.~\ref{fig:modesRs36b1}, \ref{fig:modesRs36b2} and~\ref{fig:modesRs36b4}. The positions of the DSF maxima obtained from the RPA and the ESA models is provided for comparison. The results are presented for three values of the coupling parameter, $r_s=\{16; 22; 36\}$, and three different temperatures, $\theta=\{1; 2; 4\}$. 

Let us, first, discuss the {\it plasmon} segment of the spectrum ($q \lesssim q_c$) well characterized by the DSF displayed in Fig.~\ref{fig:skdynTq1}. In this segment both ESA and 9MA dispersion relations in Figs.~\ref{fig:modesRs36b1}, \ref{fig:modesRs36b2} predict very similar positions of the plasmon resonances and exhibit a noticeable {\it redshift} with respect to the RPA result. However, as it is discussed in Sec.~\ref{Sec2C} and~\ref{SecReconstr}, at least for the densities with $r_s\gtrsim 10$, the decrement of the ESA plasmon (with the static LFC) is always underestimated (cf. Fig.~\ref{fig:Ftau}) compared to the 9MA plasmon $\Omega_1(q)$ with the dynamical correlations included via the Nevanlinna parameter function $Q_2(q,z)$. Next, with the reconstruction of the dielectric function on the complex frequency plane within the 9MA approach, we can explicitly analyze the behavior of the plasmon decrement as it approaches the pair excitation continuum (grey shaded area). The corresponding dashed red lines, $\Omega_1(q) \pm \Delta \Omega_1(q)$, in Figs.~\ref{fig:modesRs36b1}, \ref{fig:modesRs36b2}, specify the wavenumber dependence of the half-width of the lower mode $\Omega_1(q)$ (the red dots) which represents in this wavenumber segment the plasmon excitation. 

In addition, the dispersion equation, $\epsilon(q,z)=0$, predicts here a second solution $\Omega_2(q)$ (the grey dots with a solid line). In our opinion, this additional solution should be considered for these wavenumbers as a {\it virtual mode}, since its decrement is found to be comparable to the excitation energy, i.e. $\Delta\Omega_2(q) \gtrsim \Omega_2(q)$. Physically, the presence of such a solution for $q\gtrsim 0.5 q_F$ can indicate that the main plasmon mode is superimposed on a broad multi-excitation continuum with the center of mass and the characteristic half-width characterized by the resolved parameters $\{\Omega_2(q), \Delta\Omega_2(q)\}$. This interpretation applies equally in the considered spectral domain ($q\lesssim q_c$) to all density and temperature cases presented in Figs.~\ref{fig:modesRs36b1}, \ref{fig:modesRs36b2}, \ref{fig:modesRs36b4}. 
Next, a close inspection of the DSF in Fig.~\ref{fig:skdynTq1} (for $\theta \leq 4$) implies that the $\Omega_2(q)$ solution in the frequency range $1.5\lesssim \omega/\omega_p\lesssim 3$ has a significantly reduced spectral weight compared to the plasmon mode and does not lead to the DSF structure with two shifted modes.              
It is also interesting to observe that the $\Omega_2(q)$ dispersion converges to the plasmon mode $\Omega_1(q)$ for $q\sim q_c$ with $q_c\sim 1.5 q_F$, which can indicate that the physical nature of excitations changes for $q \geq q_c$.   

Indeed, for $q > q_c$ the short wavelength segment with the negative  plasmon dispersion is followed by the {\it roton}-like minimum.
Besides, exactly in this region we observed a discontinuity in the resolved dispersion relation $\Omega_1(q)$, which is preceded by the divergence of the plasmon decrement by approaching $q\sim 1.5 q_c$ (cf. the shaded red area bounded by $\Omega_1(q)\pm \Delta \Omega_1(q)$ in Fig.~\ref{fig:modesRs36b1}). A similar discontinuity but now for the visually well resolved two-mode solution is clearly observed near $q\sim 3 q_F$ ($\sim 2 q_c$). It is also preceded by the divergence of the modes' decrements $\Delta \Omega_{1(2)}$. Finally, for larger wavenumbers, $q> 3.4 q_F$, the lower mode $\Omega_1(q)$ (indicated now by blue dots) shifts closer to the position of the parabolic RPA-dispersion centered in the pair excitation continuum, while the upper mode $\Omega_2(q)$ (shown here by grey solid line with dots) possesses a very large decrement, and physically, due to the interaction effects, represents the multi-excitation contributions beyond the upper bound of the ideal Fermi gas, $\hbar \omega \gtrsim \epsilon_{q+q_F}-\epsilon_{q_F}$. 

In summary, the performed detailed analysis of the dispersion relations and the $q$-dependence of the modes' decrements permits us to clearly distinguish {\it three characteristic wavenumber segments} with quasi-excitations of different nature, confirming the physically expected result. 
First, the dispersion equation predicts the usual plasmon which is followed by the roton feature observed for $1.8 q_F\lesssim q\lesssim 3 q_F$. The {\it roton} segment is always revealed in the dispersion relation of the first mode $\Omega_{1}(q)$ and is accompanied by the higher-frequency branch $\Omega_{2}(q)$, however, only in the same roton wavenumber domain, and approaching a lower bound specified by the double plasmon excitation, $2\, \omega_p(q)$ (cf. Fig.~\ref{fig:skdynTq10}), when the density is diminished ($r_s=36$).  

A clear distinction of the transition point between the roton and single-particle segments becomes more difficult at higher temperatures and densities. At an intermediate temperature (cf. $\theta=2$ in Fig.~\ref{fig:modesRs36b2}) the roton segment is still observable when $2 q_F\lesssim q\lesssim 3 q_F$, but the lower mode $\Omega_1(q)$ is significantly damped due to the decay into particle-hole excitations. In contrast, the upper mode is not influenced by this decay channel being well above the pair excitation continuum, and has a significantly smaller decrement. The right-hand boundary of the roton-segment around $q\sim 3 q_F$ again can be identified by a steep increase in $\Delta \Omega_{1,2}(q)$,
in particular, in the strong coupling case ($r_s=36$).  

For higher densities/temperatures such that when $r_s \leq 16$ or $\theta \geq 4$ the decrement $\Delta \Omega_{1}(q)$ of the main mode is drastically enhanced and overlaps with the high-frequency solution $\Omega_2(q)$, cf. $\theta=4$ in Fig.~\ref{fig:modesRs36b4}. At these conditions both modes become virtual.

The ESA and RPA models do not describe such a complicated spectrum structure though the roton feature is seen in the unique ESA eigenmode. 

\begin{figure}[t]
\hspace{-0.9cm}
\includegraphics[width=0.515\textwidth]{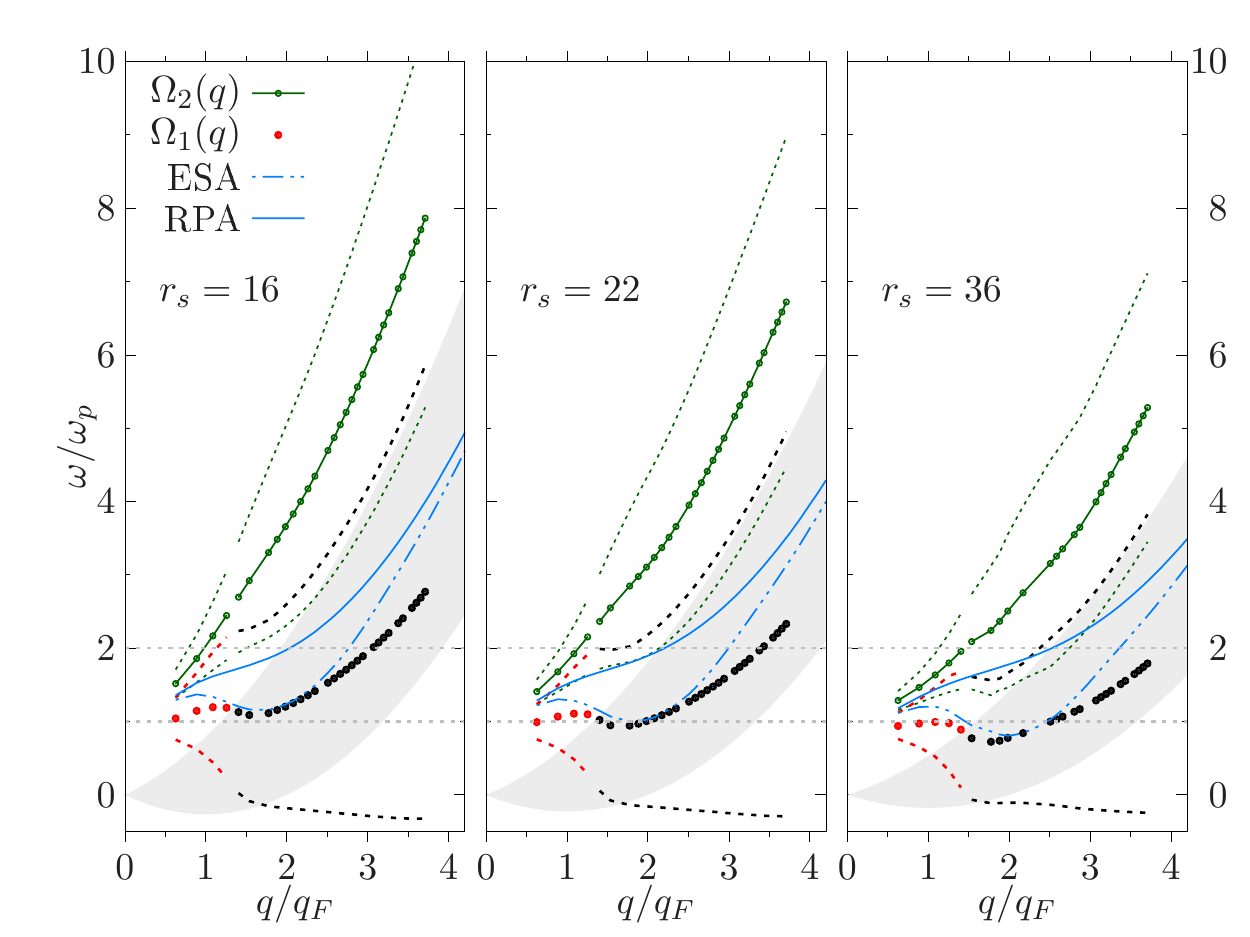}
\vspace{-0.3cm}
\caption{As in Fig.~\ref{fig:modesRs36b1} but for $\theta=4$. The frequencies $\omega^{\text{SH}}_{3(4)}(q)$ are found by the Shannon entropy maximization procedure. This choice is physically substantiated as in Fig.~\ref{fig:modesRs36b4} since both options reproduce $F(q,\tau)$ within the statistical error bars.}
\label{fig:modesRs36b4SHAN}
\end{figure}

Finally, in Fig.~\ref{fig:modesRs36b4SHAN} we demonstrate that at higher temperatures the Shannon EM approach, once used for the reconstruction of the higher characteristic frequencies $\omega_{3(4)}(q)$, does not qualitatively influence the physical results for the observed roton feature and the supplemental high-frequency shoulder due to multi-excitations as compared to the optimized solution (9MA) presented in Fig.~\ref{fig:modesRs36b4}.

A comparative discussion of the high-energy branch is provided in the next subsection.

\section{Discussion and outlook}\label{Outlook}

\subsection{Comparative discussion of the dispersion relation}

It is natural now to compare our approach to the analysis of the dynamical properties of Fermi fluids of charged particles within existing standard methods of quantum statistical physics, in particular, those based on the calculation of the Feynman diagrams. Traditionally, these calculations are reduced to the evaluation of 
the leading corrections to the RPA bubble, i.e., to the evaluation of the DLFC $G\left( q,\omega \right)$ function in a certain approximation and under certain conditions. In particular, the role of the short-range dynamical correlations in the density response of the homogeneous electron gas in the high-density limit corresponding to some simple metals was studied in detail in Refs.~\cite{diag1, diag2, diag3, diag5, diag41, diag42,  PhysRevB.31.2779, PhysRevB.31.2796, PhysRevB.31.5837}. By “short-range” we mean any physical correlation mechanism other than the collective plasma oscillation, whose macroscopic Coulomb origin is well understood through the random-phase approximation. These efforts were driven by the observation of the DSF shape presenting either a double peak or a main peak with shoulders, the shape which could not be described within the unextended RPA. First, the correlated-basis-functions theoretical method was employed whose advantage was that it provided a clear physical insight into the physical processes leading to the observable characteristics like $S(q, \omega)$ and the inverse longitudinal dielectric function $\epsilon^{-1}\left( q,\omega \right)$ interrelated by the fluctuation-dissipation theorem. Computations of the leading proper polarization Feynman diagrams outside the particle–hole continuum performed by Sturm and Gusarov~\cite{diag3} permitted to go beyond the RPA and to describe (in the high-density limit and at zero temperature) the DSF structure attributed to the correlation-induced double-plasmon excitations. Further on, an even better agreement with the observed complicated DSF structure with the second harmonic of the original plasmon excitation in a significantly broader realm of variation of density and temperature was achieved within a complete dynamic theory for the electron gas at high to metallic densities. This theory (valid for large and small momentum transfers and at high to metallic electron densities) was combining the dominant features of the shielded-interaction and the T-matrix approximations with the conservation sum rules~\cite{PhysRevB.31.2779, PhysRevB.31.2796, PhysRevB.31.5837}. It was found within this theory that the dynamic properties of the resulting polarization function and the dynamic structure factor could not be adequately approximated by the local-field constructions. In particular, the non-local effects were demonstrated to be important for the dynamic properties of the electron gas, see~\cite{PhysRevE.54.3518,PhysRevB.95.035416,PhysRevB.37.9268,PhysRevB.59.10145,dornheim-etal.nl.2020prl}. In addition, the higher harmonic generation in strongly coupled classical plasmas was earlier observed using the method of molecular dynamics and described~\cite{Hartmann_2009} in terms of the nonlinear generalization 
of the quasi-localized-charge approximation~\cite{GoldenKalman1}.

On the other hand, there is a formal non-linear algebraic relation between the DLFC and the dynamical Nevanlinna function 
$Q_{2}\left( q, z \right)$, see~\cite{PhysRevE.81.026402}, constructed here to satisfy nine sum rules and thus involving three- and four-particle correlations. 
Though the present extended self-consistent method of moments based on this Nevanlinna function is completely within the linear response theory, it has permitted us to observe a clear sign of the dynamical correlation effects in the strongly coupled UEG ($r_s\geq 16$, $\theta\sim 1$). They manifest themselves as the observed bi-modal structure of the DSF, in a finite momentum range $q_c \leq q\leq 2 q_c$, with the sharp high-frequency resonances (cf. Fig.~\ref{fig:skdynTq10}) at the position predicted by the second solution of the dispersion equation, $\Omega_2(q)$, see   Fig.~\ref{fig:modesRs36b1}.
The critical momentum $q_c$ can be estimated as the crossing point of the plasmon dispersion curve with the upper bound of the pair excitation continuum, $\omega_p\left( q_c \right)\approx q_c^2+2 q_c$. 

For the wavenumbers $q > q_c$ we can, following~\cite{Pitaevskii}, assume that a
high energy branch is formed due to the interaction of several quasiparticle excitations. To observe this multi-excitation as a distinct spectral feature, their combined energy should be above the parabolic upper bound of the pair continuum, and, the constituting quasiparticles should have a sufficiently long lifetime.  Such criteria, in the case of strongly coupled UEG can be satisfied by two possible combinations: two plasmons + roton (2P+R) or two rotons + plasmon (2R+P). Since for $r_s\geq 16$ the roton minimum lies well below the plasmon frequency, the (2P+R)-states most probably will decay into the lower energy (2R+P)-states, and only the contribution from the latter will dominate in the spectral density. Moreover, the integrated density of quasiparticles is scaled proportionally to $S(q)$ which indicates their high population near the roton minimum, and, consequently, a higher probability of the 2R-excitation over a double plasmon. This simple considerations lead us to a qualitative explanation the position of the higher-energy branch in Figs.~\ref{fig:modesRs36b1},\ref{fig:modesRs36b2}, where the second solution of the dispersion equation $\Omega_2(q)$ is seen as a combination of the 2R state,  $2\Omega_1(q)$, and a plasmon of the energy $\omega_p$.

We wish to mention here a few details more. The moment approach permits to construct an analytical expression for the dielectric function $\epsilon\left( q,\omega \right)$ and to analyze the intrinsic discrepancies between the locations of the broad peaks in the DSF spectrum and the explicit solutions of the corresponding dispersion equation. Another advantage of the method of moments is that the involved sum rules for any mathematically correct Nevanlinna function are satisfied automatically so that even in the static approximation for the latter, the emerging local field due to the intimate link with conservation principles, is still a qualitatively correct dynamic characteristic permitting to go beyond the relaxation-style modifications of the RPA similar to the Mermin theory~\cite{PhysRevE.81.026402}. On the other hand, static approximations to the local field~\cite{PhysRevLett.125.235001} only modify the static potential in the RPA and lead to no qualitative change in the shape of $S\left( q,\omega \right)$.

\subsection{Conclusions and outlook}

The predicted new shape of the UEG spectrum at low density/strong coupling ($r_s\geq 16$) of the electrons, constitutes the main result of the present work achieved, in addition, with a significantly lower computational effort and with much lower complexity in comparison to the quantum Monte-Carlo path-integral method based on the DLFC reconstruction~\cite{dornheim.2018prl,groth.2019prb}. 
The relative simplicity of the method of moments for theoretical and numerical calculations allows to carry out the on fly reconstruction of the dynamical characteristics of warm and dense uniform electron fluids of variable density and coupling.

The interrelation between the PIMC-generated dynamic local-field correction and the nine-moment Nevanlinna function is to be studied in detail elsewhere. 

Our results testify the importance of the dynamical correlation effects in terms of multiple excitations beyond the particle-hole band. The single particle-hole excitations alone are presumably not sufficient to explain the obtained $\textit {ab initio}$ results for the intermediate scattering function and the interrelated high-frequency tail of the dynamical structure factor. 

The predictions of the present results, as well as the description of the 
position and magnitude of the observed high-frequency branch, deserve, in our opinion, future experimental investigations which will provide deeper understanding of the collective excitations in the UEG in the low-density/strong coupling regime. 

The input required by the suggested approach is reduced to that of a limited set of frequency moments and the simulation data on only two static  characteristics, the static structure factor $S(q)$ and the static value of the system dielectric function. Both quantities can be accurately estimated from the first-principle PIMC simulations~\cite{dornheim.2018prl,filinov-etal.2021ctpp}. For an approximate evaluation of $S(q)$ a broad list of methods is available, e.g. the effective static local-field (ESA) parametrization~\cite{PhysRevLett.125.235001}, the hypernetted-chain method~\cite{Tanaka_HNC2016}, and the STLS scheme~\cite{1986JPSJ}.

On the other hand, the observation of new details in the system spectrum is directly related to the incorporation to the model of four higher-order sum rules not taken into account in earlier models. The values of these sum rules are determined here using the Shannon entropy maximization procedure optimized, where necessary, by the PIMC calculations of the ISF. Their direct determination in terms of the three- and four-particle static correlation functions found using the PIMC approach could be a difficult but interesting work to do. In one-component electron liquids, but not, e.g., in hydrogen-like two-component plasmas~\cite{PerelEliashberg, PhysRevE.62.5648, PhysRevE.64.056410, PhysRevE.90.053102, PhysRevE.91.019903}, even more frequency moments/sum rules converge and it might be curious to investigate their influence on the eigenmodes. The SCMM permits to carry out such a development but it remains to be seen whether it would lead to observable new details of the system dynamical properties.   

The obtained algebraic expressions for the inverse dielectric function and other dynamical quantities can be also employed in a variety of WDM applications and beyond, e.g. in the interpretation of XRTS experiments~\cite{Thomson_scattering}, analysis of the ion stopping power models~\cite{Cayzac.2017,Zhen-Guo.2017} or to evaluate the ionization potential depression in dense plasmas~\cite{Ionization_potential,zan-etal.2021pre}.

\section*{Acknowledgments}
This article is dedicated to the memory of A.N. Starostin, V.E. Fortov, and E.E. Son. 
The authors acknowledge the support by the Deutsche Forschungsgemeinschaft via Project No. BO1366-15 and the grant \#AP09260349 of the
Ministry of Education and Science, Kazakhstan. IMT is grateful to V.M. Adamyan from the Odessa National University
(Ukraine) for introducing him to the method of moments and for
fruitful collaboration.

\section*{Appendix: Construction of dynamical Nevanlinna function}

Some mathematical aspects of the moment approach are
provided along the mathematical details of the nine- and five-moment
versions of the self-consistent method of moments. 

\textit{Frequency power moments.} With the spectral density chosen as the loss function, see Eq.~(\ref{lossf1}) in the main text, it stems from the detailed-balance condition, 
\begin{equation}
S\left( q,-\omega \right) =\exp \left( -\beta \hbar \omega \right) S\left(
q,\omega \right) \ ,
\end{equation}%
that 
\begin{equation}
C_{\nu }\left( q\right) =\frac{4\pi ne^{2}}{\hbar q^{2}}\left[ 1+\left(
-1\right) ^{\nu }\right] \mu_{\nu-1} \ ,
\end{equation}%
where
\begin{equation}
\mu_{\nu} =\int_{-\infty }^{\infty }\omega
^{\nu}S\left( q,\omega \right) d\omega \ ,\quad \nu=-1,1,3,5,7\ 
\end{equation}%
are the moments of the dynamic structure factor. 
The fact that we account for the vanishing moments $\left\{
C_{\nu }\left( q\right) =0\right\} $, $\nu =1,3,5,7$, 
is reflected in the relatively simple five- and nine-moment forms of the
Nevanlinna formula \cite{krein-book} employed in the main text and besides we have
that 
\begin{equation}
\omega _{j}^{2}\left( q\right) =\frac{\mu_{2j-1}}{\mu_{2j-3}}\ ,\quad
j=1,2,3,4.
\end{equation}%
Due to the Cauchy-Schwarz-Bunyakovsky inequalities, the
conditions 
\begin{eqnarray}
0<\omega_{1}\left( q\right) <\omega_{2}\left( q\right) <\omega_{3}\left( q\right) <\omega_{4}\left( q\right)
\label{CSB}
\end{eqnarray}
should be satisfied to warrant the fulfillment of the required mathematical properties
of the Nevanlinna and the inverse dielectric functions.

To mention that Nevanlinna's theorem can be proven on the basis of
the technique of generalized resolvents of M.G. Krein, see \cite{sb1,sb2}.
Further details of the method of moments can be found in \cite{Varentsov_2005}.

\textit{The dynamical Nevanlinna function.}
Nevanlinna's formula \cite{shohat-book} establishes a one-to-one linear-fractional
transformation between all solutions of the Hamburger problem and all
Nevanlinna functions $Q_{n}\left( q,z\right) $ such that\textit{\ }$%
\lim_{z\rightarrow \infty }Q_{n}\left( q,z\right) /z=0$:%
\begin{eqnarray}
\int_{-\infty }^{\infty }\frac{d\mathcal{L}\left( q,\omega \right) }{%
z-\omega }=\frac{E_{n+1}\left( z;q\right) +Q_{n}\left( q,z\right)
E_{n}\left( z;q\right) }{D_{n+1}\left( z;q\right) +Q_{n}\left( q,z\right)
D_{n}\left( z;q\right) }\ ,\nonumber \\
\quad n=0,1,2,\ldots \quad  \label{eq:Nevf}
\end{eqnarray}%
The coefficients of this transformation are polynomials $%
D_{n}\left( z;q\right) $ orthogonal with the weight $\mathcal{L}\left( q,\omega \right) 
$, which can be easily constructed using the standard Gram-Schmidt procedure,
while the polynomials $E_{n}\left( z;q\right) $ are their conjugate \cite%
{tkachenko-book}. In the main text we consider the five- and nine-moment Hamburger
problems so that we need only the following polynomials:%
\begin{eqnarray}
&&D_{2}\left( z;q\right) =\left( z^{2}-\omega _{1}^{2}\right),\; 
D_{3}\left( z;q\right) =z\left( z^{2}-\omega _{2}^{2}\right),\nonumber\\ 
&&D_{2}\left( z;q\right) =\left( z^{2}-\omega _{1}^{2}\right),\; 
D_{3}\left( z;q\right) =z\left( z^{2}-\omega _{2}^{2}\right),\nonumber\\ 
&&E_{2}\left( z;q\right) =C_{0}z\ ,\; E_{3}\left( z;q\right)
=C_{0}\left( z^{2}-\left[ \omega _{2}^{2}-\omega _{1}^{2}\right] \right),\nonumber\\
&&E_{4}\left( z;q\right) =C_{0}\left( z^{3}+b_{1}z\right),\; E_{5}\left(
z;q\right) =C_{0}\left( z^{4}+d_{2}z^{2}+d_{0}\right).\nonumber \\
\label{DE}
\end{eqnarray}
Here, 
\begin{eqnarray*}
&&b_{1}=\frac{\omega _{1}^{4}-2\omega _{1}^{2}\omega _{2}^{2}+\omega
_{2}^{2}\omega _{3}^{2}}{\omega _{1}^{2}-\omega _{2}^{2}}\ ,\\
&&d_{2}=\frac{\omega _{1}^{2}\left( \omega _{2}^{2}-\omega _{3}^{2}\right) +\omega
_{3}^{2}\left( \omega _{4}^{2}-\omega _{2}^{2}\right) }{\omega
_{2}^{2}-\omega _{3}^{2}}\ , \\ 
&&d_{0}=\omega _{1}^{2}\omega _{2}^{2}+\omega _{3}^{2}\frac{\omega
_{1}^{2}\left( \omega _{4}^{2}-\omega _{2}^{2}\right) +\omega _{2}^{2}\left(
\omega _{3}^{2}-\omega _{4}^{2}\right) }{\omega _{2}^{2}-\omega _{3}^{2}}\ .%
\end{eqnarray*}
In the case of $5=2n+1$ moments, by virtue of the Kramers-Kronig relations, we
arrive at the expression for the inverse dielectric function provided in Eq.~(\ref{idf}) in the main text. In quantum systems we abandon the static approximation for the Nevanlinna function
\begin{equation}
h_{2}\left( q\right) =Q_{2}\left( q,0\right) =\omega _{2}^{2}\left( q\right)
/\left( \sqrt{2}\omega _{1}\left( q\right) \right)\ ,   \label{h2}
\end{equation}%
and reconstruct the dynamic five-moment Nevanlinna function by equalizing
the r.h.s. of Eq.~(\ref{eq:Nevf}) with $n=2$ to the same with $n=4$: 
\begin{equation}
\frac{E_{3}+Q_{2}E_{2}}{D_{3}+Q_{2}D_{2}}=\frac{E_{5}+Q_{4}E_{4}}{%
D_{5}+Q_{4}D_{4}}\ ,  \label{25}
\end{equation}%
where from we express the five-moment Nevanlinna function in terms of the
nine-moment one:%
\begin{equation}
Q_{2}=-\frac{D_{3}E_{5}-E_{3}D_{5}+\left( D_{3}E_{4}-D_{4}E_{3}\right) Q_{4}%
}{D_{2}E_{5}-E_{2}D_{5}+\left( D_{2}E_{4}-E_{2}D_{4}\right) Q_{4}}\ .
\end{equation}%
Then, we applied to the loss function, which is obviously proportional to
the imaginary part of the r.h.s of Eq.~(\ref{25}), the procedure employed in 
\cite{arkhipov-etal.2017prl,arkhipov-etal.2020pre} to determine the five-moment-parameter (\ref{h2}), and
obtained the zero-frequency value of the nine-moment Nevanlinna function:%
\begin{equation}
Q_{4}\left( q,0\right) =ih_{4}\left( q,\tilde{\omega}\right) =\frac{i\omega
_{3}^{2}\left( \omega _{2}^{2}-\omega _{1}^{2}\right) \left( \omega
_{4}^{2}-\omega _{3}^{2}\right) }{\omega _{1}\sqrt{2\left( \omega
_{3}^{2}-\omega _{2}^{2}\right) ^{3}\left( \omega _{3}^{2}-\omega
_{1}^{2}\right) }}\ .
\end{equation}%
This approximation turned to be sufficient not only for the reliable analytical
description of the UEG-DSF QMC data, but for the direct observation of the two-mode structure of the system spectrum. Moreover, the above nine-moment
expressions simplify into the previous five-moment solution~(\ref{h2}) as soon as we consider two successive limiting transitions: $\omega_{4}\left( q\right) \rightarrow
\infty $ and $\omega_{3}\left( q\right) \rightarrow \infty$.

\bibliography{ref}

\end{document}